\documentclass[journal]{IEEEtran}

\usepackage[T1]{fontenc}
\usepackage[utf8]{inputenc}

\usepackage{amsmath,amssymb}
\usepackage{graphicx}
\graphicspath{{figures/}}
\usepackage{booktabs}
\usepackage{array}

\usepackage{cite}

\usepackage{url}
\usepackage[hidelinks]{hyperref}

\title{Assured AI-Native Network Control Loops: State of the Art, Research Challenges and the Missing Runtime Assurance Layer}

\author{Bartosz~Belter and Mariusz~G\l{}ąbowski%
\thanks{Bartosz Belter is with the Poznan Supercomputing and Networking Center, Poznan, Poland.}%
\thanks{Mariusz G\l{}abowski is with Poznan University of Technology, Poznan, Poland.}%
}

\begin{document}

\maketitle

\begin{abstract}
The evolution towards autonomous and AI-native telecommunication networks is transforming network control from predefined automation towards increasingly distributed and intelligent decision-making. Advances in closed-loop automation, Open Radio Access Network (O-RAN), Network Digital Twins, AI-driven orchestration, and autonomous agents enable multiple specialized control functions to operate concurrently across network domains and timescales. This evolution introduces a system-level assurance
challenge: decisions that are individually acceptable may interact through shared resources and network state, while changes in their operational context may invalidate the assumptions under which they were previously evaluated.

This paper presents a state-of-the-art review of AI-native network control with a particular focus on the composition and runtime assurance of autonomous control loops. It examines developments in closed-loop and zero-touch automation, intelligent network controllers, Network Digital Twins, AI-driven orchestration, autonomous agents, trustworthy AI, and runtime assurance. The analysis shows that these research directions provide important foundations for autonomous network operation, but do not
provide a unified mechanism for assuring heterogeneous network control loops whose decisions depend on shared and dynamically changing network state.

Based on this gap, the paper identifies dependency-aware runtime assurance as a research direction for assured composition of AI-native network control loops. The proposed perspective associates autonomous decisions with the assumptions and dependencies on which their validity relies, monitors changes that may invalidate previously accepted decisions, and supports runtime resolution of interactions between concurrent control actions. A telecom use case and an initial architecture and formal model are introduced to illustrate this concept and to identify open challenges in dependency representation, runtime validation, conflict resolution, latency-aware assurance, and experimental evaluation.

The resulting research direction shifts the assurance problem from validating individual AI components towards maintaining the validity and compatibility of autonomous decisions as their shared operational environment evolves.
\end{abstract}

\section{Introduction}

Telecommunication networks are entering a period of profound transformation. The evolution towards 5G-Advanced and future 6G systems is not only about higher capacity, lower latency, or new radio technologies; it changes how networks are designed, operated, and controlled \cite{DangEtAl2020}. 

Cloud-native platforms, virtualization, heterogeneous access, and increasingly dynamic service requirements combine into an infrastructure that requires substantially more flexible and sophisticated management than the comparatively monolithic architectures of previous network generations \cite{BonatiEtAl2020}.

For decades, network management relied on a stable operational model: human experts, supported by Operational Support Systems (OSS), monitored behaviour, analysed incidents, and applied configuration changes through predefined procedures. This model was workable while network environments were comparatively static and operational changes occurred at manageable timescales. In modern programmable and service-driven networks, however, increasing heterogeneity, service diversity, and the need for rapid adaptation make traditional human-centred and manually driven management increasingly inadequate \cite{SousaEtAl2021,CoronadoEtAl2022}.

Autonomous operation emerged as a response. The autonomic computing paradigm envisioned systems capable of self-configuration, self-healing, self-optimization, and self-protection with limited human intervention \cite{KephartChess2003}, and was later adapted to communication networks as autonomic communications \cite{DobsonEtAl2006,SterrittEtAl2005}. In telecommunications, this
evolved from simple automation towards closed-loop control: systems that continuously monitor the network state, analyse available information, decide on corrective actions, and apply changes. The MAPE-K model (Monitor, Analyse, Plan, Execute over a shared Knowledge base) became one of the most influential conceptual foundations for such adaptive systems \cite{KephartChess2003}, and remains a reference model for contemporary closed-loop automation frameworks \cite{BrunEtAl2009}.

Software-Defined Networking (SDN) and Network Function Virtualization (NFV) provided the technological substrate. SDN separated control from forwarding, enabling software-based, logically centralized decision-making \cite{KreutzEtAl2015}. NFV extended this
flexibility by deploying network functions as software components on virtualized infrastructure, at the cost of new challenges in
orchestration, lifecycle management, and resource optimization \cite{MijumbiEtAl2016}. The rapid progress of
Artificial Intelligence (AI) and Machine Learning (ML) added a further dimension: modern networks generate enormous operational data, and AI techniques can extract patterns that elude traditional analytical methods, with applications in traffic prediction, anomaly detection, resource allocation, and wireless optimization
\cite{ZhangEtAl2019}.

These threads converge in the vision of AI-native networks, where AI is not an external optimization tool but an integral element of network control  \cite{ZhangEtAl2025AINative}. AI components observe conditions, generate decisions, and directly influence operational behaviour. The opportunity is substantial---faster adaptation, better resource efficiency, reduced operational complexity---but so is the shift in the nature of the problem. Traditional automation rested on explicit rules and deterministic behaviour. AI-driven systems decide from learned representations, training data, and operational context, in environments that keep changing \cite{SeshiaEtAl2022}.

The difficulty becomes acute when future networks are viewed not as systems governed by a single intelligent entity but as ecosystems of autonomous components. A radio optimization function, a cloud resource manager, an energy optimization mechanism, and a service orchestration component may all operate autonomously while influencing the same underlying infrastructure. Each may behave correctly according to its own objective, yet their combined behaviour may not produce the expected system outcome. A controller optimizing throughput may raise resource consumption; another optimizing energy efficiency may reduce available
resources; both decisions can be locally reasonable while their interaction degrades overall behaviour. This phenomenon - well studied in multi-agent systems \cite{JenningsEtAl1998} - is only beginning to be recognized in telecom contexts
\cite{ErdolEtAl2026}.

This leads to a fundamental research question: how can autonomous network control loops be composed so that their joint operation remains predictable and trustworthy?

Existing research addresses many individual aspects. O-RAN introduces intelligent controllers and open interfaces for AI-driven radio optimization \cite{O-RANArchitecture2023,PoleseEtAl2023}. Network Digital Twins provide models for prediction and simulation \cite{MasaracchiaEtAl2022}. AIOps applies AI and machine learning techniques to operational monitoring, analysis, and automation. Trustworthy AI research addresses robustness, transparency, and reliability \cite{LiEtAl2023,GawlikowskiEtAl2023}.

Runtime assurance, originally developed for safety-critical autonomous systems, offers a relevant perspective. The Simplex architecture \cite{ShaEtAl2001} introduced the principle of pairing an advanced controller with a simpler safety controller capable of taking over when unsafe behaviour is detected. Seshia et al. extended this perspective towards formal verification of AI systems, arguing that safety cannot rely exclusively on design-time verification because autonomous systems
operate in complex and changing environments \cite{SeshiaEtAl2022}. More recent work has applied runtime assurance specifically to learning-enabled systems \cite{PhanEtAl2020}. Applying these ideas to telecom requires addressing additional complexity: distributed architectures, multiple control domains, and dynamic dependencies between autonomous components.

This paper presents a state-of-the-art analysis of AI-native network control with a particular focus on the missing assurance layer required for future autonomous telecom systems. It reviews the evolution from traditional management towards AI-driven control loops, analyses current AI-native architectures (O-RAN, Network Digital Twins, AI-based orchestration), and discusses the applicability of trustworthy AI and runtime assurance. The analysis identifies a research gap: current approaches provide increasingly powerful mechanisms for autonomous decision generation, but lack comprehensive methods for assuring the safe composition of multiple AI-native control loops operating in dynamic environments.

The remainder of the paper is organized as follows. Section~2 discusses the evolution towards AI-native control. Section~3 analyses current AI-native architectures and their limitations. Section~4 reviews trustworthy AI and runtime assurance.
Section~5 focuses on the composition of multiple AI-based control loops. Section~6 presents a telecom use case illustrating the dependency-aware assurance problem. Section~7 identifies the remaining research gaps. Section~8 introduces an initial architecture and formal model for assured composition. Section~9 outlines future research directions, and Section~10 concludes.

\section{Evolution Towards AI-Native Network Control}

\subsection{From Traditional Network Management to Autonomous Operation}

Traditional telecommunication network management was built around human
operators and dedicated management systems. Operational Support Systems
(OSS) provided fault, configuration, and performance management
capabilities that proved sufficient while infrastructure changes were
predictable and service requirements evolved gradually. Modern networks
no longer fit that profile: multiple access technologies, cloud-native
platforms, virtualized network functions, and highly dynamic services
combine into environments where manual intervention and predefined
procedures are increasingly difficult to maintain
\cite{CoronadoEtAl2022}.

Autonomic computing offered an early answer. Kephart and Chess described
systems capable of managing themselves through continuous observation
and adaptation---self-configuration, self-healing, self-optimization,
and self-protection \cite{KephartChess2003}. The concept was
adapted to communication networks as autonomic communications, in which
networks adapt their behaviour to changing operational conditions
\cite{DobsonEtAl2006}, with Sterritt et al. further elaborating
the self-* properties and their operational implications
\cite{SterrittEtAl2005}. The principles remain relevant; what has
changed is the scale. The number of interacting components, the
diversity of services, and the speed of operational changes in future
networks far exceed what early autonomic systems were designed for
\cite{CoronadoEtAl2022}.

\subsection{The Rise of Closed-Loop Network Automation}

The shift from static configuration to closed-loop automation was a decisive step. Rather than executing isolated management actions, closed-loop systems observe state, analyse information, determine actions, and apply changes based on feedback. The MAPE-K model (Monitor, Analyse, Plan, Execute over a shared Knowledge base) provided the influential conceptual template, separating monitoring, decision-making, and execution \cite{KephartChess2003}. In contemporary 5G and 6G networks, these principles have evolved towards increasingly autonomous and zero-touch management, combining closed-loop operation with data-driven and AI-enabled decision-making \cite{CoronadoEtAl2022}.

Software-Defined Networking supplied the architectural foundation by
separating the control plane from the forwarding plane and enabling
programmable, logically centralized control
\cite{KreutzEtAl2015}. Network Function Virtualization extended
this transformation, implementing network functions as software on
virtualized infrastructure---gaining flexibility at the cost of new
challenges in orchestration and resource allocation
\cite{MijumbiEtAl2016}. Yet early automation remained
largely rules-based: systems could execute complex procedures, but the
reasoning behind them was still designed by humans.

\subsection{From Automation Towards Intent-Based and Zero-Touch Management}

As complexity grew, it became clear that configuration-driven automation
would not suffice. Intent-Based Networking (IBN) shifted the focus from
configuring devices to declaring desired outcomes, letting the management
system determine how to achieve them \cite{RFC9315}. Mendiola et al. trace
the evolution of software-defined traffic engineering and the increasing
separation between management objectives and their realization in the
underlying network \cite{MendiolaEtAl2017}. More generally, intent-based
networking introduces a lifecycle in which high-level objectives are
translated into network actions and continuously assessed against the
intended outcome \cite{RFC9315}. This creates an additional assurance
problem: independently specified intents may interact through shared
resources and network state, requiring mechanisms for detecting and
resolving conflicting objectives.

In parallel, the ETSI Zero-touch Network and Service Management (ZSM)
framework defined architectural principles for highly automated operation
through closed-loop automation, hierarchical management, and increased
autonomy \cite{ETSI_ZSM002}. Oladele et al. provide a comprehensive
survey of ZSM, identifying the lack of cross-domain assurance mechanisms
as a primary barrier to achieving full autonomy
\cite{LiyanageEtAl2022}. The ETSI Experiential Networked Intelligence
(ENI) framework complements ZSM by defining an AI-enabled system
architecture that applies context-aware, intent-based policies
\cite{ETSI_ENI_005}, and the ITU-T Y.3172 recommendation specifies
an architectural framework for ML in future networks including IMT-2020
\cite{ITU_T_Y3172}.

Both approaches raise a fundamental question: how much autonomy can be
safely delegated? Moving from automation to autonomy means the system
must interpret conditions, select actions, and adapt to situations not
explicitly anticipated at design time.

\subsection{AI as an Enabler of Adaptive Network Control}

AI and Machine Learning created new possibilities where
traditional automation reached its limits. Modern networks
generate vast amounts of operational data---performance measurements,
alarms, configuration data, and traffic observations---from which
learning-based methods can extract patterns that are difficult to
capture with manually designed models. Zhang et al. surveyed deep
learning applications across mobile and wireless networking, including
traffic prediction, resource allocation, mobility management, and
network optimization \cite{ZhangEtAl2019}. AIOps extends the use of
AI to operational processes such as failure detection, anomaly
analysis, root-cause analysis, and remediation
\cite{ZhangEtAl2025AIOps}. Reinforcement-learning approaches further
extend this paradigm from prediction towards sequential decision-making,
with applications including dynamic network access, routing, resource
sharing, and resource management \cite{LuongEtAl2019}.

In this role, AI served mainly as an enhancement of existing
processes: models provided predictions and recommendations,
while humans or deterministic frameworks retained final decision
authority. The benefits were real, but the role of AI was still
supporting, not governing.

\subsection{Towards AI-Native Network Control Loops}
AI-native networking goes further. AI becomes part of the
control architecture itself: models observe conditions, generate
decisions, and influence operational behaviour within closed
loops. Saad et al. identified AI as a key enabling technology for
future wireless systems, supporting autonomous operation and
adaptive resource management \cite{SaadEtAl2020}. Letaief et al.
articulated the roadmap towards AI-empowered wireless networks,
positioning AI as an increasingly integral element of future
network design \cite{LetaiefEtAl2024}. More recent work on
AI-native 6G architectures takes this integration further by
embedding AI capabilities directly into network components and
combining distributed AI data and computing functions with
collaborative network control \cite{ZhangEtAl2025AINative}.

The benefits are faster adaptation, improved efficiency, and reduced
operational complexity. The cost is a new source of uncertainty:
decisions come from learned representations of the environment rather
than from explicit rules. The challenge sharpens when future networks
are viewed as distributed ecosystems of specialized AI components---radio
optimization, transport management, cloud orchestration, service
assurance---operating simultaneously across different domains. Building
intelligent controllers is no longer the only question. Ensuring that
multiple intelligent controllers can operate together without
unpredictable interactions is the open problem---and the motivation for
a runtime assurance perspective.

\section{AI-Native Network Control Architectures}

\subsection{Architectural Foundations of AI-Native Networks}

Making AI integral to network control requires more than inserting
machine learning models into existing management systems. AI needs access
to operational data, the ability to influence behaviour, and mechanisms
to execute decisions within complex infrastructure. Several
architectural directions address this: O-RAN introduces programmability
and AI-driven optimization into radio access; Network Digital Twins
provide virtual representations for prediction and experimentation;
AI-driven orchestration extends automation across domains; and recent AI
agents bring autonomous reasoning into operational systems. These
approaches have largely evolved independently, each focused on specific
challenges. The result is a landscape rich in individual solutions but
without a unified perspective on how multiple autonomous components
should interact safely within a shared environment \cite{ErdolEtAl2026}.

\subsection{O-RAN and Intelligent Network Controllers}

The Open Radio Access Network (O-RAN) architecture is
one of the most visible examples of embedding intelligence
directly into telecom control. Where traditional RAN relied on
vendor-specific, closed control mechanisms, O-RAN introduces
open interfaces, virtualization, and programmable intelligence
\cite{PoleseEtAl2023}. Its central elements include the RAN
Intelligent Controllers (RICs): the Non-Real-Time RIC supports
longer-term optimization, policy management, and AI/ML
lifecycle functions, while the Near-Real-Time RIC enables
faster control through specialized applications (xApps)
\cite{PoleseEtAl2023}. Bonati et al. describe the broader open,
programmable, and virtualized mobile-network ecosystem on
which these developments build, illustrating the transition
from functions embedded in dedicated network elements towards
software-based network control \cite{BonatiEtAl2020}.
The flexibility cuts both ways. Multiple xApps may operate
simultaneously in overlapping network areas while pursuing
different objectives. Because independently developed xApps
can affect common network parameters and performance metrics,
their actions may result in direct or indirect conflicts and
degrade network performance \cite{ErdolEtAl2026}. Existing
O-RAN conflict-mitigation mechanisms are predominantly
rule-based and may resolve conflicts by suppressing selected
actions or rolling back configurations after performance
degradation is detected \cite{ErdolEtAl2026}. At the same time,
the openness, virtualization, and programmability introduced
by O-RAN expand the system's attack surface and create
additional security challenges for its interfaces, platforms,
and intelligent control functions \cite{LiyanageEtAl2023ORANSecurity}.
O-RAN therefore provides mechanisms for deploying and
executing intelligent control functions, but deployment
capability does not by itself guarantee that decisions produced
by independently developed autonomous functions remain
compatible at the system level.

\subsection{Network Digital Twins as a Foundation for Predictive Control}

A Network Digital Twin (NDT) is a dynamic virtual representation
of a physical network that is maintained using information from
its physical counterpart and can support monitoring, analysis,
simulation, prediction, and optimization. Originating from the
broader Digital Twin paradigm developed for cyber-physical and
industrial systems, the concept has increasingly been applied to
telecommunication networks. Masaracchia et al. survey Digital
Twin applications for 6G and identify synchronization, data
management, modelling, and the integration of physical and
virtual environments among the important challenges for their
practical deployment \cite{MasaracchiaEtAl2022}.

More recent work positions NDTs as an enabling technology for
network management and control. Raza et al. describe NDTs as
virtual network representations that collect information from
physical, virtual, and software components and support performance
analysis, emulation, and intelligent control
\cite{RazaEtAl2025}. In the RAN context, Vilà et al. demonstrate
how an NDT can maintain a representation of network conditions
and support the evaluation of management and optimization
strategies \cite{VilaEtAl2023}.

For autonomous networks, the attraction is clear: an AI controller can
evaluate possible actions within a virtual representation before
executing them on a production network, testing configuration changes,
predicting performance impacts, and reducing operational risk. A Digital
Twin can answer:

\begin{quote}
``What is likely to happen if this action is executed?''
\end{quote}

It does not, on its own, answer:

\begin{quote}
``Should this action be allowed to execute under the current operational
conditions?''
\end{quote}

That distinction---between prediction and assurance---matters
increasingly as networks move towards autonomous operation.

\subsection{AI-Driven Orchestration Across Network Domains}

Future services depend on interactions across radio access,
transport, edge computing, and cloud infrastructures.
Traditional orchestration handles lifecycle management,
including service deployment, resource allocation, and
coordination across heterogeneous infrastructure. AI-driven
orchestration extends these capabilities with prediction,
optimization, and adaptive decision-making. The ETSI ZSM
framework defines architectural principles for highly automated
management through closed-loop automation, management
domains, and increased autonomy \cite{ETSI_ZSM002}, while
the ETSI ENI framework specifies an AI-enabled system
architecture that uses contextual information and policies to
support adaptive network management \cite{ETSI_ENI_005}.
Together, these developments illustrate the evolution from
conventional service orchestration towards increasingly
autonomous management spanning multiple network and
computing domains.

Cross-domain autonomy exposes a problem that grows with intelligence.
Different orchestration components optimize different parts of the
system: a cloud orchestrator may minimize computational cost while a
network controller minimizes latency. Each decision may be individually
reasonable; their interaction may degrade overall service behaviour.
Future orchestration therefore needs mechanisms that consider not only
individual decisions but also relationships and dependencies between
autonomous components.

\subsection{AI Agents and Autonomous Network Operations}

Large language models and autonomous AI agents open another direction.
Unlike ML models trained for specific optimization tasks, AI agents
combine perception, reasoning, planning, and interaction with external
tools. Wang et al. survey LLM-based autonomous agents, proposing a
profiling-memory-planning-action framework that characterizes their
architecture \cite{WangEtAl2023}. Xi et al. extend this with a
comprehensive taxonomy of agent capabilities, identifying tool use,
memory management, and multi-agent collaboration as key enablers
\cite{XiEtAl2025}. Sumers et al. propose a cognitive architecture
for language agents, drawing on cognitive science to formalize the
perception-reasoning-action loop \cite{SumersEtAl2024}.

In networking, Liu et al. survey the application of large
language models to network operations and management,
covering network design, automation, optimization, and
security, as well as their integration with paradigms such
as SDN, NFV, intent-based networking, and zero-touch
management \cite{LiuEtAl2025LLMNetOps}. Their analysis
also identifies explainability, computational scalability,
privacy, legacy-system integration, and domain knowledge
as important barriers to operational deployment
\cite{LiuEtAl2025LLMNetOps}. More recent work extends
this direction towards LLM-based multi-agent systems for
network management, where autonomous agents assume
roles such as coordination, translation, and negotiation,
while reliability, hallucination, and inference latency remain
important challenges \cite{KimEtAl2026LLMMAS}.
Agents introduce challenges distinct from traditional
optimization models. Their behaviour depends on generated
plans, context interpretation, and interactions with external
tools, making it difficult to guarantee that decisions remain
within expected operational boundaries. The more flexible
the decision-making, the more important mechanisms that
can evaluate and constrain it become.

\subsection{Limitations of Current AI-Native Architectures}

The building blocks are present: O-RAN enables intelligent control
applications, Digital Twins provide predictive capabilities,
orchestration frameworks enable cross-domain automation, and AI agents
introduce advanced reasoning. What they address is how to generate
better autonomous decisions. What they provide significantly less
support for is how to assure that multiple autonomous decisions remain
compatible when executed together. Several questions remain unresolved:

\begin{itemize}
    \item how to detect conflicts between independently operating AI controllers,
    \item how to verify whether assumptions behind an AI decision remain valid,
    \item how to represent dependencies between autonomous components,
    \item how to maintain predictable behaviour in a continuously changing network environment.
\end{itemize}

Future AI-native networks therefore require an additional architectural
capability: intelligence for autonomous decision generation, paired with
assurance that makes that autonomy operationally trustworthy.

\section{Runtime Assurance and Trustworthy AI for Autonomous Networks}

AI-native control changes how network decisions are generated. In
traditional management, decisions come from deterministic algorithms,
explicit policies, and human-defined procedures; even when automated,
the logic is predictable and analysable before deployment. AI-based
control derives decisions from learned representations of complex
environments. This enables optimization where analytical models are
incomplete, but introduces new uncertainty: behaviour depends on
training data, model assumptions, operational context, and interactions
with the surrounding system. In telecommunications, where control
decisions influence service availability, resource utilization, and
stability, a model that performs well in a laboratory does not
automatically behave safely in a dynamic production network. The
question shifts from whether an AI model performs well to whether an
AI-driven control system can be trusted during operation.

This section reviews trustworthy AI and runtime assurance approaches and
their relevance for future autonomous network control. The objective is
not a general overview of AI safety, but to identify which concepts
support the operation of multiple interacting AI-based control loops in
complex telecom environments.

\subsection{From AI Performance to AI Trustworthiness}

Early ML adoption in network management focused on optimization
performance---prediction accuracy, classification performance, resource
utilization, convergence speed. These metrics remain important but give
only a partial view. A model may excel under training and validation
conditions yet produce unreliable results when the operational
environment shifts, as it does continuously in telecom: traffic
patterns, user behaviour, topology, service requirements, and
infrastructure conditions all evolve.

Trustworthy AI research responds to this gap by addressing broader
properties---robustness, transparency, explainability, accountability,
privacy, fairness. The European Commission's High-Level Expert Group
emphasized that AI systems should be technically capable and aligned
with human values, legal requirements, and societal expectations,
defining seven requirements spanning human agency, technical robustness,
privacy, transparency, diversity, societal wellbeing, and accountability
\cite{HLEGTrustworthyAI2019}. NIST's AI Risk Management Framework
defines trustworthiness characteristics including validity, reliability,
safety, security, transparency, explainability, and fairness, structured
around four functions---govern, map, measure, manage---that collectively
address AI risk throughout the system lifecycle
\cite{NISTAIRMF2023}. Li et al. survey the transition from
principles to practices in trustworthy AI, identifying a persistent gap
between framework-level guidelines and operational implementation,
particularly for distributed, multi-component systems
\cite{LiEtAl2023}. Amodei et al. framed the problem as concrete
technical challenges---reward hacking, side effects, scalable oversight,
safe exploration---that remain relevant for AI-driven network control
\cite{AmodeiEtAl2016}, and Brundage et al. analysed the malicious
use of AI, highlighting that distributed autonomous systems introduce
attack surfaces not present in traditional deployments
\cite{BrundageEtAl2018}.

These frameworks are valuable, but they are formulated primarily from
the perspective of AI applications and models. Autonomous network
control adds a dimension. The AI component is not isolated; it is
embedded in a feedback loop where its actions modify the environment
that shapes future decisions. Trustworthiness must therefore be
considered at the level of the complete control system---including
interactions between AI models, network state, operational policies, and
other autonomous components. Ensuring that an individual model is robust
does not ensure that a system of multiple AI-driven controllers behaves
safely. A controller may produce a valid decision by its own objective
while creating undesirable effects for another controller in the same
infrastructure.

\subsection{Limitations of Model-Centric Trustworthy AI Approaches}

Much of trustworthy AI research focuses on individual models.
Explainable AI methods make behaviour more understandable by providing
insights into the relationship between inputs and outputs: Guidotti et
al. survey methods for explaining black-box models, categorizing them
into model-agnostic and model-specific approaches, and noting that
explanations generated post-hoc do not guarantee that the model itself
behaves safely in novel conditions \cite{GuidottiEtAl2018}.
Uncertainty-aware learning quantifies confidence in predictions:
Gawlikowski et al. provide a comprehensive survey of uncertainty in
deep neural networks, distinguishing between epistemic uncertainty
(model knowledge gaps) and aleatoric uncertainty (data noise), and show
that uncertainty quantification improves decision robustness but does
not address interactions between multiple models operating in the same
environment \cite{GawlikowskiEtAl2023}. Adversarial robustness
techniques harden models against perturbations; Hendrycks et al. provide
an overview of catastrophic AI risks, arguing that robustness must be
treated as a system-level property rather than a model-level
optimization target \cite{HendrycksEtAl2022}.

These approaches address uncertainty within the model. They do not
capture uncertainty originating from the operational environment or from
interactions with other autonomous components.

Consider two AI controllers in a future network. The first optimizes
energy consumption by reducing active resources during low demand. The
second optimizes latency by increasing resource availability in
anticipation of traffic growth. Both may operate correctly by their own
objectives and available information. The system-level outcome may still
be undesirable if their assumptions about the environment are
incompatible. Improving individual model accuracy does not solve this;
it requires mechanisms that understand the operational context in which
decisions execute and continuously evaluate whether the conditions
supporting a decision remain valid. Autonomous networks therefore
require a transition from model-centric trustworthiness towards
system-level assurance: not whether an AI model is reliable, but whether
an autonomous control action remains acceptable given the current system
state, dependencies, and actions of other components.

\subsection{Runtime Assurance in Safety-Critical Autonomous Systems}

Runtime assurance has been investigated in safety-critical autonomous
systems where complex decision-making components operate in environments
that cannot be fully predicted in advance. The central idea is to
separate advanced, potentially uncertain decision-making from an
independent assurance mechanism responsible for maintaining safe
operation. The Simplex architecture \cite{ShaEtAl2001} combines an
advanced controller with a simpler safety controller that takes over
when unsafe behaviour is detected, originally addressing systems in
which high-performance controllers could not themselves provide the
required guarantees.

Seshia et al. extend the assurance discussion to learning-enabled
systems, arguing that verification of AI-based components requires
reasoning about their interaction with complex and changing
environments rather than treating the learned model in isolation
\cite{SeshiaEtAl2022}. Runtime assurance architectures
subsequently adapted the Simplex principle to learning-enabled
controllers. Phan et al. propose the Neural Simplex Architecture,
combining a neural-network-based advanced controller with a verified
baseline controller and a decision module that switches control when
safety can no longer be guaranteed \cite{PhanEtAl2020}.

The same principle has also been extended beyond a single autonomous
agent. Mehmood et al. introduce a Distributed Simplex Architecture for
multi-agent systems, in which each agent executes a local assurance
mechanism while system-level safety follows from the interaction of
these local guarantees \cite{MehmoodEtAl2021}. This demonstrates that
runtime assurance can be generalized to distributed autonomous control,
while also exposing a substantially harder problem: safety must then be
reasoned about not only for individual decisions but also for
interactions among independently operating controllers.

For autonomous telecommunications, this offers an attractive direction.
Instead of requiring every AI model to be completely predictable---which
may be unrealistic for highly adaptive systems---a dedicated assurance
layer could continuously evaluate whether AI-generated actions remain
within acceptable operational boundaries.

\subsection{Applying Runtime Assurance Concepts to Network Control}

The principles transfer, but not directly. Telecommunication networks
differ from many traditional safety-critical systems: they are highly
distributed, continuously evolving, and composed of multiple interacting
domains. In an autonomous vehicle, the assurance mechanism typically
evaluates a single system operating within a physical environment. In a
future telecom network, multiple autonomous components are distributed
across infrastructure layers---radio optimization functions, transport
controllers, cloud resource managers, service orchestration systems,
operational AI agents---each operating on different objectives, data
sources, and timescales. A radio controller may optimize spectrum
efficiency within milliseconds; an orchestration system may optimize
service placement over minutes or hours. Overall network behaviour
emerges from the interaction of these partially independent processes.

This distributed nature creates challenges that traditional runtime
assurance does not address. The problem is no longer whether one
autonomous component behaves safely, but whether the combined behaviour
of multiple components remains aligned with global objectives. An action
acceptable in isolation may become problematic after another controller
modifies the environment or invalidates the assumptions on which the
original decision rested.

This points to a distinction between design-time and runtime
assurance. Design-time approaches establish confidence based on
models, requirements, and assumptions available before deployment.
For self-adaptive systems, however, some of these assumptions may
change or become invalid during operation. Weyns et al. introduce
the concept of perpetual assurances, arguing that assurance for
self-adaptive systems must be an enduring process spanning the
system's operational lifetime and continuously incorporating new
evidence under uncertainty \cite{WeynsEtAl2019}.
Calinescu et al. operationalize a related idea through the ENTRUST
methodology, which combines design-time and runtime modelling and
verification with dynamically maintained assurance cases. The
approach is intended to provide evidence that a self-adaptive system
continues to satisfy its requirements while its environment,
architecture, or operating conditions change
\cite{CalinescuEtAl2018}.

For telecom networks, assurance mechanisms should continuously evaluate
not only the outcome of AI decisions but also the validity of the
conditions under which they were generated---network state, available
resources, active policies, dependencies between services, and actions
performed by other autonomous entities. AI-based controllers may
discover relationships that are difficult to represent explicitly, so
assurance must combine operational monitoring with contextual knowledge
about reasoning assumptions and dependencies. Runtime assurance should
not be a safety mechanism applied after decisions are generated, but an
integral part of the control architecture, enabling autonomous
components to operate while maintaining awareness of their operational
boundaries.

\subsection{Dependency-Aware Assurance for Composed AI Control Loops}

Most existing assurance mechanisms consider autonomous components as
individual entities. Future AI-native networks will be characterized by
composition of multiple control loops, and network control functions are
inherently interdependent: changing radio resource allocation influences
transport requirements; modifying cloud resource placement affects
service latency; applying energy optimization strategies influences
resilience. Assurance cannot rest on evaluating individual actions. It
must consider dependencies between actions, system states, and control
objectives. A decision is valid only within the context in which its
underlying assumptions remain satisfied.

This introduces dependency-aware runtime assurance. Every autonomous
decision should be associated with explicit or implicit assumptions
describing the conditions required for safe execution:

\begin{itemize}
    \item expected network state and resource availability,
    \item validity of input data and observations,
    \item absence of conflicting actions from other controllers,
    \item compliance with operational policies and service requirements,
    \item stability of dependencies between network components.
\end{itemize}

When these assumptions change, a previously approved decision may no
longer hold. The assurance mechanism should continuously monitor
relevant dependencies and determine whether autonomous actions remain
acceptable. This differs from conventional AI trustworthiness, which
focuses on characteristics of the model itself---accuracy, robustness,
explainability. Dependency-aware assurance focuses on the relationship
between AI decisions and the evolving system in which they execute. For
future 6G networks, where autonomy emerges not from a single intelligent
entity but from the interaction of many specialized AI components,
without mechanisms for assuring their composition, increasing autonomy
may unintentionally introduce new forms of operational instability
\cite{ErdolEtAl2026}.

The key research challenge is therefore not only how to create
intelligent network controllers, but how to enable multiple intelligent
controllers to operate together predictably and trustworthy. This
requires a runtime assurance layer capable of understanding
dependencies, validating decision context, and supporting safe
adaptation when operational assumptions change. Existing AI-native
network architectures provide powerful mechanisms for autonomous
optimization; they lack systematic approaches for assuring the safe
composition of multiple AI-driven control loops.

\section{Composition of Autonomous Network Control Loops}

The building blocks for autonomous operation already exist:
intelligent controllers, Network Digital Twins, AI-driven
orchestration, and autonomous agents. These technologies enable
increasingly sophisticated autonomous functions, but future
network operation is unlikely to rely on a single centralized
intelligence. Instead, control is increasingly distributed among
specialized functions operating across different domains,
objectives, and timescales. O-RAN provides a concrete example
of this evolution, with multiple control functions operating
through different control loops and potentially acting on
overlapping network resources
\cite{PoleseEtAl2023,ErdolEtAl2026}. The challenge therefore
shifts from building individual intelligent controllers to
ensuring that independently operating control functions can
safely coexist.

\subsection{From Individual Control Loops to Composed Autonomous Systems}

Closed-loop control has traditionally been considered within
a single optimization objective: a controller observes state,
selects an action, and receives feedback. This works for
individual functions---mobility optimization, fault management,
or resource allocation---implemented as separate loops. The
situation changes when multiple loops operate simultaneously
within the same infrastructure. Each may be locally correct
and individually optimized, yet their interaction may produce
unexpected system-level effects. Jennings et al. highlighted
that coordination between autonomous agents is a fundamental
challenge in multi-agent systems because individual behaviours
influence the overall outcome \cite{JenningsEtAl1998}.
A closely related problem appears in multi-agent reinforcement
learning. Gronauer and Diepold identify non-stationarity as one
of the fundamental challenges of multi-agent learning: because
multiple agents adapt and act within a shared environment, the
environment observed by an individual agent changes as the
policies of other agents change \cite{GronauerDiepold2022}.
Zhang et al. review the theoretical foundations of multi-agent
reinforcement learning and show that extending results from
single-agent reinforcement learning to multi-agent settings
introduces fundamentally different theoretical challenges,
including convergence and learning in stochastic games
\cite{ZhangYangBasar2021}.

Consider several AI-driven controllers operating at once: a radio
optimization controller improving spectral efficiency, an energy
management controller reducing power consumption, a service
orchestration controller optimizing application placement, and a fault
management agent adapting configuration after detecting anomalies. Each
pursues a valid objective. The problem appears when their actions
interact. Increasing radio capacity may improve user experience but
raise energy consumption. Moving workloads closer to users may reduce
latency but create additional transport requirements. A fault
mitigation may restore one service while affecting another. The
system-level outcome emerges from composition, not from any single
decision analysed in isolation.

\subsection{Challenges of Multi-Objective Autonomous Control}

Composing autonomous loops is difficult because objectives are rarely
independent. Traditional optimization assumes a clearly defined
objective function---minimize latency, maximize throughput, reduce
energy. In real environments, network objectives conflict: performance
versus energy efficiency, reliability versus resource utilization,
automation speed versus operational safety, local optimization versus
global objectives. Multi-objective optimization provides mathematical
tools for such trade-offs \cite{BertsekasTsitsiklis1989}, but
autonomous operation adds a further dimension: different controllers may
not share the same optimization criteria or even the same understanding
of system state. A controller optimizing one objective may
unintentionally modify the conditions under which another operates, so
that a previously valid decision becomes inappropriate after another
autonomous entity changes the environment.

The coexistence of multiple autonomous control functions
introduces an explicit coordination problem. In O-RAN,
independently developed xApps may operate over overlapping
network resources and pursue different optimization objectives,
leading to direct and indirect conflicts that can degrade
network performance \cite{ErdolEtAl2026}. Similar concerns
arise in intent-based management, where multiple intents may
express objectives that interact or conflict and therefore
require mechanisms for conflict detection and resolution
\cite{RFC9315}. These observations show that correctness of
an individual control decision cannot be considered independently
of the other control functions and system conditions with which
it interacts.

This creates a temporal dimension of assurance. It is not
enough to verify that a decision is acceptable when generated.
Such a decision is evaluated under assumptions about the
current system state, available resources, active policies, and
the behaviour of other control functions. If these conditions
change before the decision is executed, the basis on which the
assurance decision was made may no longer hold. Assurance
must therefore address not only whether a decision is acceptable,
but also how long that assessment remains valid and which
changes in its dependencies should trigger re-evaluation.

\subsection{Interactions Between Control Loops and Emergent Behaviour}

Complex behaviour may emerge from relatively simple interactions between
components---a property widely studied in distributed systems,
multi-agent environments, and cyber-physical systems
\cite{Sycara1998}. In autonomous networks, emergence may occur when
multiple controllers respond simultaneously to changing conditions. A
sudden traffic increase may trigger a traffic engineering controller to
modify routing, a cloud orchestrator to allocate resources, and a radio
controller to adjust capacity. Each action changes the environment
observed by the others. Without coordination and dependency awareness,
autonomous loops may enter undesirable states: oscillations, inefficient
resource allocation, or continuous corrective actions against each
other.

The problem is compounded because AI-based controllers may not expose
explicit decision logic. Traditional conflict detection relies on
predefined rules or known dependencies. ML models and AI agents generate
decisions from complex internal representations that are difficult to
analyse before execution \cite{GuidottiEtAl2018,LiEtAl2023}. Future
autonomous networks therefore need mechanisms capable of reasoning about
interactions between loops, not only evaluating individual actions.

\subsection{Limitations of Current Approaches}

Existing frameworks provide important mechanisms for implementing
autonomous network functions, but they address their interactions
to different degrees. O-RAN, for example, enables multiple intelligent
applications to influence RAN behaviour, while concurrent xApps
pursuing different objectives may generate conflicting control actions
and degrade network performance
\cite{ErdolEtAl2026}. Intent-based and zero-touch
management frameworks provide mechanisms for coordinating automated
network operation across management domains, but their primary focus
is on expressing, realizing, and assuring management objectives rather
than on providing a general runtime assurance mechanism for arbitrary
compositions of autonomous controllers
\cite{RFC9315,CoronadoEtAl2022}. Network Digital Twins provide
complementary capabilities for modelling, simulation, prediction,
and evaluation of network behaviour, while their ability to represent
the operational network depends on maintaining sufficient fidelity
and synchronization with the physical system
\cite{MasaracchiaEtAl2022,RazaEtAl2025}. Trustworthy AI, in turn,
provides principles and techniques for increasing confidence in
individual AI components, including properties such as robustness,
reliability, and explainability \cite{LiEtAl2023}.
Taken together, these approaches address important parts of the
autonomous-network-control problem, but none provides a general
mechanism for continuously assuring the composition of independently
operating control loops as their shared dependencies and operational
conditions evolve. This motivates an assurance mechanism that operates
across individual control loops and evaluates not only their individual
decisions, but also the validity of those decisions in the context of
their interactions and changing shared environment.

\subsection{Towards Assured Composition of AI-Native Control Loops}

Assured composition treats the network not as a collection of
independent optimization components but as a system of interacting
autonomous decision-making entities. It requires answering questions
currently insufficiently addressed:

\begin{itemize}
    \item Are the assumptions behind an AI decision still valid?
    \item Has another autonomous component changed the operational context?
    \item Do multiple planned actions create conflicting effects?
    \item Can a decision be executed safely under current conditions?
\end{itemize}

Answering these requires combining several capabilities:

\begin{itemize}
    \item continuous monitoring of network state and dependencies,
    \item representation of relationships between autonomous components,
    \item validation of decisions against operational constraints,
    \item mechanisms for preventing or mitigating unsafe interactions.
\end{itemize}

Autonomous networks require not only intelligent controllers but an
assurance layer responsible for maintaining safe and predictable
behaviour when multiple AI-driven control loops operate simultaneously.
The challenge is not replacing autonomous decision-making, but making
autonomy operationally trustworthy.

\section{Research Gap and Open Challenges}

The previous sections show significant progress towards
autonomous and AI-native network operation. Intelligent
controllers enable adaptive and increasingly AI-driven network
optimization \cite{PoleseEtAl2023,ErdolEtAl2026}; Network
Digital Twins provide modelling, simulation, prediction, and
evaluation capabilities \cite{MasaracchiaEtAl2022,RazaEtAl2025};
intent-based and zero-touch management frameworks enable
increasingly automated operation across management domains
\cite{RFC9315,CoronadoEtAl2022}; and advances in AI,
reinforcement learning, and autonomous agents extend network
control from prediction towards increasingly sophisticated
decision-making \cite{LuongEtAl2019,LiuEtAl2025LLMNetOps}.
Taken individually, these developments address important
elements of autonomous network operation. Taken together,
however, they expose a different problem. Future networks are
likely to comprise multiple independently developed control
functions operating concurrently over shared resources, network
state, and infrastructure. Evidence from programmable network
architectures already shows that individually reasonable control
actions can interact and produce conflicts or degraded system
performance \cite{ErdolEtAl2026}. Consequently, increasing the
capability of individual controllers is not sufficient: assurance
must also consider the interactions among controllers and the
conditions under which their decisions remain valid.
The research gap addressed in this work is therefore the lack
of a general assurance perspective for the composition of
autonomous network control loops under changing operational
conditions. In particular, existing approaches do not provide
a unified mechanism for representing the dependencies on which
assurance decisions rely, detecting when changes in those
dependencies invalidate previously accepted decisions, and
coordinating the resulting control actions while maintaining
system-level operational constraints.

\subsection{Beyond AI-Enabled Optimization Towards Assured Autonomy}

Much of existing research in AI-enabled networking focuses on
optimization performance---throughput, latency, energy efficiency,
prediction accuracy \cite{ZhangEtAl2019}. These
contributions are valuable but address only one dimension. Autonomous
networks require decisions that are not only effective but appropriate
given the current operational context: whether the assumptions behind a
decision remain valid, and whether other autonomous actions may affect
its outcome.

Consider two scenarios. In the first, an AI controller predicts a
traffic increase and allocates additional resources; the prediction is
correct and no other component modifies the environment, so the
decision improves performance. In the second, another autonomous
controller simultaneously performs energy optimization and reduces
available resources. Both decisions are individually reasonable; their
combination may degrade service quality. The problem is not a lack of
intelligence in individual components but the absence of mechanisms
capable of assuring their interaction. This suggests a transition from
AI-enabled optimization towards assured autonomy---mechanisms for
generating decisions, and mechanisms for continuously evaluating whether
those decisions remain valid and safe.

\subsection{Limitations of Existing Architectural Approaches}

Table~\ref{tab:research_gap} summarizes the capabilities and limitations identified
across the main research directions considered in this survey.

\begin{table*}[t]
\centering
\caption{Research gaps in current AI-native network approaches}
\label{tab:research_gap}

\begin{tabular}{p{0.20\textwidth}p{0.28\textwidth}p{0.40\textwidth}}
\toprule
\textbf{Approach} &
\textbf{Primary contribution} &
\textbf{Remaining challenge} \\
\midrule

AIOps &
Operational analytics, anomaly detection, prediction &
Limited mechanisms for assuring autonomous actions generated from operational insights \\

O-RAN Intelligent Controllers &
Programmable AI-driven RAN optimization &
Limited support for coordination and assurance between multiple intelligent applications \\

Network Digital Twins &
Simulation, prediction, and what-if analysis &
Dependence on model accuracy and limited runtime validation of autonomous decisions \\

AI Agents &
Flexible reasoning and autonomous task execution &
Uncertain behaviour and limited guarantees in dynamic operational environments \\

Trustworthy AI &
Model robustness, explainability, risk management &
Primarily focused on individual AI systems rather than composed autonomous systems \\

Runtime Assurance &
Safe operation of autonomous systems &
Limited adaptation to distributed multi-controller telecom environments \\

\bottomrule
\end{tabular}

\end{table*}

The comparison shows that existing approaches address
complementary parts of the autonomous network control problem,
but provide different levels of support for interactions among
multiple autonomous control functions. AIOps and machine
learning for networking improve operational intelligence,
prediction, anomaly analysis, and optimization
\cite{ZhangEtAl2019,ZhangEtAl2025AIOps}. O-RAN provides
an architecture for deploying multiple intelligent control
applications, while concurrent xApps can introduce direct and
indirect conflicts requiring explicit coordination mechanisms
\cite{PoleseEtAl2023,ErdolEtAl2026}. Network Digital Twins
provide modelling, simulation, prediction, and evaluation
capabilities, with their effectiveness depending on model
fidelity and synchronization with the operational network
\cite{MasaracchiaEtAl2022,RazaEtAl2025}. LLM-based agents
introduce increasingly flexible reasoning, planning, and tool
interaction, while reliability, explainability, and predictable
operational behaviour remain important challenges
\cite{WangEtAl2023,LiuEtAl2025LLMNetOps}. Trustworthy AI
primarily addresses properties of AI systems and models,
including robustness, reliability, transparency, and
explainability \cite{LiEtAl2023,NISTAIRMF2023}. Runtime
assurance provides mechanisms for maintaining safety despite
uncertain or learning-enabled controllers and has also been
extended to distributed multi-agent systems
\cite{PhanEtAl2020,MehmoodEtAl2021}.
Taken together, however, these research directions do not
provide a unified approach for continuously assuring the
composition of heterogeneous network control loops whose
decisions depend on shared and dynamically changing network
state. In particular, the comparison exposes limited support
for explicitly representing decision dependencies, tracking
the validity of assurance results as those dependencies change,
and coordinating concurrent autonomous actions under
system-level operational constraints.

The comparison highlights that existing approaches address important
parts of the challenge but provide only partial support for assuring
interactions between multiple autonomous control loops. AIOps and ML
for networking improve operational intelligence, prediction, anomaly
analysis, and optimization
\cite{ZhangEtAl2019,ZhangEtAl2025AIOps}. O-RAN enables multiple
intelligent control applications, while concurrent xApps may generate
direct and indirect conflicts that require explicit coordination
\cite{PoleseEtAl2023,ErdolEtAl2026}. Network Digital Twins provide
modelling, simulation, prediction, and evaluation capabilities, with
their effectiveness depending on model fidelity and synchronization
with the operational network
\cite{MasaracchiaEtAl2022,RazaEtAl2025}. AI agents introduce flexible
reasoning, planning, memory, and tool interaction, while reliability
and predictable operational behaviour remain important deployment
challenges
\cite{WangEtAl2023,XiEtAl2025,LiuEtAl2025LLMNetOps}. Trustworthy AI
addresses properties such as robustness, reliability, transparency,
and explainability of AI systems
\cite{LiEtAl2023}. Runtime assurance provides mechanisms for maintaining
safe operation despite uncertain or learning-enabled controllers and
has also been extended to distributed multi-agent systems
\cite{PhanEtAl2020,MehmoodEtAl2021}.

Taken together, however, these approaches do not provide a unified
mechanism for continuously assuring the composition of heterogeneous
network control loops whose decisions depend on shared and dynamically
changing network state.

\subsection{Open Research Challenge 1: Runtime Validation of AI Decisions}

Traditional software systems can often be analysed based on explicit
logic and well-defined requirements. AI-based controllers introduce
additional uncertainty because their behaviour depends on learned
models, operational data, and changing environmental conditions
\cite{SeshiaEtAl2022,GawlikowskiEtAl2023}.
Runtime assurance provides an important foundation for addressing this
problem. The Simplex architecture separates an advanced controller from
a simpler safety controller that can take over when safe operation can
no longer be guaranteed \cite{ShaEtAl2001}. This principle has been
extended to learning-enabled controllers through the Neural Simplex
Architecture \cite{PhanEtAl2020} and to distributed multi-agent systems
through the Distributed Simplex Architecture
\cite{MehmoodEtAl2021}. These approaches demonstrate that runtime
assurance can protect systems containing complex or distributed
autonomous controllers.

The remaining challenge considered here is more specific: autonomous
telecommunication networks may contain heterogeneous control loops
developed independently, operating at different timescales and over
shared network state and resources. Assurance in such an environment
must account for the dependencies on which individual control decisions
were evaluated and determine when changes in those dependencies require
previously accepted decisions to be re-evaluated.

\begin{itemize}
    \item How can the assumptions behind an AI-generated decision be represented?
    \item How can their validity be continuously monitored?
    \item How should the system react when assumptions become invalid?
\end{itemize}

A particularly difficult aspect is the acquisition of $\alpha_i$ from
black-box models such as deep neural networks and LLM-based agents, whose
internal representations are not directly interpretable
\cite{GuidottiEtAl2018,LiEtAl2023}. Section~\ref{sec:assumption_acquisition}
addresses this by proposing a multi-source approach in which $\alpha_i$
is constructed from exogenous constraints, decision-context capture,
XAI-based approximation, and runtime profiling rather than extracted
from the model's internals. Nevertheless, the completeness and accuracy
of $\alpha_i$ under this approach remain open research questions,
particularly for models whose behaviour is highly non-linear or
context-dependent.

These questions remain insufficiently addressed in current AI-native
network architectures.

\subsection{Open Research Challenge 2: Dependency-Aware Control Loop Composition}

Future networks are expected to comprise multiple specialized
autonomous control functions operating at different abstraction levels
and timescales. These functions may interact through shared network
state, resources, and infrastructure. O-RAN already provides a concrete
example of such interactions: independently developed xApps may affect
overlapping network parameters and performance objectives, resulting in
direct or indirect conflicts \cite{PoleseEtAl2023,ErdolEtAl2026}.

An important challenge is therefore to make the dependencies relevant
to autonomous decisions explicit. A controller may evaluate an action
based on assumptions about resources, network state, policies, or other
operational conditions that can subsequently be changed by another
control function. A dependency-aware assurance mechanism should provide
a representation of relationships between:

\begin{itemize}
    \item control loops,
    \item managed resources,
    \item operational objectives,
    \item decision assumptions,
    \item external constraints.

\end{itemize}

Such a representation would enable the assurance mechanism to determine
which operational conditions a decision depends on and to evaluate
whether changes in those conditions affect the continued validity of
that decision. It would also provide a basis for identifying interactions
between concurrently operating autonomous control functions.

\subsection{Open Research Challenge 3: Safe Adaptation in Dynamic Environments}

Telecommunication networks are inherently dynamic: new services appear,
traffic patterns change, failures occur, and configurations evolve.
Consequently, assurance cannot be treated solely as a design-time or
one-time activity. Research on self-adaptive systems has similarly
argued for assurance as an enduring process that continuously
incorporates new evidence as the system and its environment evolve
\cite{WeynsEtAl2019}. Dynamic assurance approaches further demonstrate
how runtime modelling and verification can be combined with evolving
assurance evidence to maintain confidence in system behaviour under
changing operating conditions \cite{CalinescuEtAl2018}.

Applying this principle to autonomous network control introduces several
challenges:

\begin{itemize}
    \item maintaining an up-to-date representation of system dependencies,
    \item handling uncertainty in operational observations,
    \item balancing autonomous operation with intervention mechanisms,
    \item defining appropriate fallback strategies.
\end{itemize}

Excessive restrictions may limit the benefits of autonomy, whereas
insufficient restrictions may expose the network to unacceptable
operational risks.

A further dimension of this challenge concerns the performance overhead
introduced by assurance itself. Control functions may operate under
strict timing constraints; for example, the O-RAN Near-RT RIC operates
on control timescales between 10~ms and 1~s
\cite{PoleseEtAl2023}. An assurance mechanism inserted into such a
control path must therefore complete the required evaluation without
violating the timing constraints of the corresponding control loop.
This creates a trade-off between assurance depth and computational
latency: how much validation can be performed within the available time
budget, and how should the assurance process adapt when that budget is
constrained? Section~\ref{sec:latency_budget} analyses this trade-off
in the context of the proposed architecture.

\subsection{Open Research Challenge 4: Conflict Resolution and Priority Arbitration}

Detecting a conflict is necessary but not sufficient. Once the assurance
layer determines that $Valid(d_i, t) = false$ or that two actions
conflict, it must decide what to do: modify, delay, reject, or roll
back. This decision involves priority arbitration---determining which
controller's objective prevails---and requires criteria that go beyond
the conflict detection itself.

Section~\ref{sec:resolution} outlines an initial resolution framework
based on service criticality, policy precedence, temporal urgency,
reversibility, and severity of impact. Several questions remain open:

\begin{itemize}
    \item How should priority be assigned when controllers operate in
    different domains with incomparable objectives (e.g., radio
    optimization vs.\ cloud orchestration)?
    \item How should the system handle conflicts where no clear priority
    ordering exists---i.e., both controllers are equally critical?
    \item What are efficient reversal procedures for non-trivially
    reversible actions, and how should the system classify actions by
    reversibility at decision time?
    \item Under what conditions is negotiated resolution between
    controllers feasible and beneficial compared to unilateral
    resolution by the assurance layer?
\end{itemize}

These questions are closely related to the broader problem of
coordination among autonomous agents. Multi-agent systems research has
long recognized that independently operating agents must coordinate
their actions when their decisions interact or when they depend on
shared resources \cite{JenningsEtAl1998}. Distributed runtime assurance
further demonstrates that safety mechanisms can be decentralized across
multiple interacting autonomous agents \cite{MehmoodEtAl2021}.
The challenge considered here is more specific: conflict resolution
must be integrated into the runtime assurance of heterogeneous
telecommunication control loops, where controllers may pursue
incomparable objectives, operate at different timescales, and affect
shared network resources. Resolution decisions must therefore account
not only for whether actions conflict, but also for their operational
priority, potential impact, reversibility, and the time available for
intervention.

\subsection{Towards a Research Framework for Assured AI-Native Networks}

The identified challenges indicate a new research direction focused on
assured composition of AI-native network control loops. Such a framework
should not replace existing AI-based optimization but complement it with
an assurance capability operating across autonomous components
\cite{ITU_T_Y3172}. Its core
capabilities include:

\begin{itemize}
    \item modelling dependencies between autonomous control loops,
    \item validating the operational context of AI-generated actions,
    \item monitoring changes that invalidate previous decisions,
    \item supporting safe execution, adaptation, or rejection of autonomous actions.
\end{itemize}

Assurance becomes an active component of the autonomous control
architecture rather than a final verification step. Future AI-native
networks require mechanisms that enable not only intelligent
decision-making but also trustworthy composition of multiple autonomous
control loops operating within a shared and evolving infrastructure.

\section{Research Directions Towards Assured AI-Native Network Control}

The evolution towards autonomous networks is progressing rapidly.
AI-based optimization, intelligent controllers, digital representations
of network infrastructures, and automated orchestration provide
important foundations for increasingly autonomous operation
\cite{LetaiefEtAl2024,ZhangEtAl2025AINative}. As autonomy increases,
network control is also becoming more distributed, with multiple
specialized control functions operating across different domains,
objectives, and timescales. O-RAN provides a concrete example of this
evolution, where multiple intelligent applications may concurrently
influence shared network resources and performance objectives
\cite{PoleseEtAl2023,ErdolEtAl2026}.

Generating effective decisions will therefore no longer be sufficient.
Future networks will also require mechanisms that ensure that autonomous
decisions remain valid, mutually compatible, and operationally safe as
the shared network environment changes.

\subsection{From Intelligent Control Towards Assured Autonomous Control}

The trajectory of network automation can be described as a transition
through several stages. The first focused on operational efficiency
through the automation of repetitive management tasks. The second
introduced closed-loop mechanisms that monitor network conditions and
execute adaptation procedures, progressively evolving towards
zero-touch network and service management
\cite{CoronadoEtAl2022,LiyanageEtAl2022}. The third brought
AI-based optimization, enabling systems to learn patterns from
operational data and increasingly support or generate network control
decisions \cite{ZhangEtAl2019,LuongEtAl2019}. The next stage considered
in this work is autonomous control under assurance constraints.

In this model, intelligence and assurance are complementary. AI
components provide increasingly adaptive decision-making capabilities,
while assurance mechanisms evaluate whether the resulting decisions
remain acceptable as the system and its operational environment evolve.
This perspective is consistent with the concept of perpetual assurance,
which treats assurance as an enduring process spanning the operational
lifetime of a self-adaptive system \cite{WeynsEtAl2019}. Runtime
assurance architectures provide a complementary foundation by
separating advanced decision-making from mechanisms responsible for
maintaining safe operation, including in learning-enabled and
distributed autonomous systems
\cite{ShaEtAl2001,PhanEtAl2020,MehmoodEtAl2021}.

For future telecommunication networks, the objective is therefore not
to restrict autonomy, but to make increasing autonomy compatible with
predictable and acceptable system-level behaviour. Traditional
management approaches face increasing difficulty in coping with the
complexity and dynamic requirements of highly automated networks
\cite{LiyanageEtAl2022}, while autonomous decision-making without
appropriate assurance may introduce interactions and consequences that
cannot be assessed solely at the level of individual controllers.
Future network control therefore requires a balance between
adaptability and assurance.

\subsection{Dependency-Aware Runtime Assurance as a Research Direction}

A central research direction emerging from this survey is the development
of dependency-aware runtime assurance mechanisms for AI-native network
control. Runtime assurance provides established mechanisms for protecting
systems that contain complex or learning-enabled controllers
\cite{ShaEtAl2001,PhanEtAl2020}, and these principles have also been
extended to distributed multi-agent systems
\cite{MehmoodEtAl2021}. In parallel, programmable network architectures
such as O-RAN demonstrate that multiple independently developed control
functions may concurrently influence shared network resources and
performance objectives, creating direct and indirect conflicts
\cite{PoleseEtAl2023,ErdolEtAl2026}.

The challenge considered here is to connect these perspectives by making
the dependencies underlying assurance decisions explicit. A
dependency-aware assurance framework should address:

\begin{itemize}
    \item What assumptions support a specific AI-generated decision?
    \item Which network components, resources, and control loops depend
    on these assumptions?
    \item How can changes in the operational environment invalidate
    previously accepted decisions?
    \item How can conflicting actions between autonomous controllers be
    identified before execution?
\end{itemize}

Addressing these questions requires combining concepts from several
research areas:

\begin{itemize}
    \item trustworthy artificial intelligence
    \cite{LiEtAl2023,NISTAIRMF2023},
    \item runtime verification and runtime assurance
    \cite{WeynsEtAl2019,PhanEtAl2020},
    \item verification of learning-enabled autonomous systems
    \cite{SeshiaEtAl2022},
    \item multi-agent coordination and distributed assurance
    \cite{JenningsEtAl1998,MehmoodEtAl2021},
    \item autonomous and zero-touch network management
    \cite{CoronadoEtAl2022,ETSI_ZSM002}.
\end{itemize}

The research challenge lies not in applying these concepts independently,
but in integrating them into an operational assurance framework for
heterogeneous telecommunication control loops. Such a framework must
reason about dependencies between decisions and network state, preserve
the validity of assurance decisions as that state evolves, and operate
within the timing constraints imposed by network control loops.

\subsection{Towards Composable Autonomous Network Control}

Traditional network architectures rely on well-defined interfaces,
protocols, and abstractions to enable interoperability between network
components \cite{KreutzEtAl2015}. AI-native networks introduce an
additional challenge that may be described as behavioural
interoperability: autonomous components may correctly exchange
information through standardized interfaces while still producing
control decisions that interact in undesirable ways. O-RAN provides a
concrete example, where independently developed xApps can operate through
a common architecture and interfaces yet generate direct or indirect
conflicts when influencing overlapping network parameters and objectives
\cite{PoleseEtAl2023,ErdolEtAl2026}.

Future autonomous network architectures should therefore consider not
only whether control components can communicate, but also whether their
decisions can be safely composed. Supporting such composition requires
representations that capture relevant properties of autonomous control,
including:

\begin{itemize}
    \item objectives of autonomous controllers,
    \item operational constraints and policies,
    \item dependencies between decisions, network state, and resources,
    \item uncertainty associated with observations and decisions
    \cite{GawlikowskiEtAl2023},
    \item potential interactions with other control loops.
\end{itemize}

Such representations could provide a basis for coordination mechanisms
in which autonomous components or an independent assurance mechanism
validate, adapt, delay, or reject decisions according to the broader
network context. Intent-based networking provides a related foundation
by separating desired outcomes from their realization and by defining
intent fulfillment and intent assurance as part of the intent lifecycle
\cite{RFC9315}. Extending this perspective from the assurance of
individual intents towards the runtime assurance of interacting control
loops represents a further research direction for composable autonomous
network control.

\subsection{Experimental Evaluation and Benchmarking Challenges}

Current AI-based networking research commonly evaluates individual
algorithms with respect to specific prediction or optimization
objectives \cite{ZhangEtAl2019,LuongEtAl2019}. While necessary, such
evaluation does not fully capture the complexity introduced when
multiple autonomous control functions operate concurrently within the
same network environment. Future experimental methodologies should
therefore be capable of assessing:

\begin{itemize}
    \item interactions between multiple autonomous control loops,
    \item robustness under changing operational conditions,
    \item impact of conflicting optimization objectives
    \cite{ErdolEtAl2026},
    \item effectiveness of runtime assurance mechanisms
    \cite{PhanEtAl2020,MehmoodEtAl2021},
    \item trade-offs between autonomy, assurance, and operational
    performance.
\end{itemize}

Such evaluation requires experimental environments capable of combining
programmable network infrastructure, autonomous control functions,
orchestration mechanisms, and realistic operational data. Open and
programmable mobile-network platforms provide foundations for deploying
and experimentally evaluating intelligent control applications
\cite{BonatiEtAl2020,PoleseEtAl2023}, while Network Digital Twins
provide complementary capabilities for modelling, simulation, and
evaluation of network behaviour under controlled conditions
\cite{MasaracchiaEtAl2022,VilaEtAl2023}. Reproducible experimental
methodologies will be increasingly important for comparing approaches
and evaluating progress towards assured autonomous network control.

A specific evaluation priority is the measurement of the latency
overhead introduced by runtime assurance. As discussed in
Section~\ref{sec:latency_budget}, assurance mechanisms must operate
within the timing constraints of the control loops they protect.
For example, the O-RAN Near-RT RIC operates on control timescales
between approximately 10~ms and 1~s \cite{PoleseEtAl2023}.
Experimental evaluation should therefore quantify the processing time
introduced by individual assurance operations, including state
monitoring, assumption validation, conflict detection, and enforcement,
under varying workloads and dependency-graph sizes. The evaluation
should distinguish synchronous assurance operations that delay decision
execution from asynchronous mechanisms that can operate outside the
critical control path. Such measurements can determine which assurance
operations are feasible at different control timescales and characterize
the trade-off between assurance depth and latency.

\subsection{Research Outlook}

Achieving practical network autonomy requires addressing challenges
that extend beyond the performance and optimization capabilities of
individual AI components. The evolution towards increasingly
AI-empowered and autonomous network operation
\cite{LetaiefEtAl2024,ZhangEtAl2025AINative} shifts attention from
the capabilities of individual intelligent functions towards the
behaviour of systems in which multiple autonomous control functions
operate concurrently.

The central research question emerging from this survey is:

\begin{quote}
How can multiple AI-native network control loops be composed and operated
in a way that preserves operational objectives, respects system
dependencies, and maintains trustworthy behaviour under changing network
conditions?
\end{quote}

Answering this question requires a shift from isolated AI optimization
towards system-level assurance of autonomous network behaviour. Future
research should investigate architectures, assurance mechanisms, and
experimental methodologies that enable multiple autonomous control
functions to operate over shared network state and resources while
maintaining acceptable system-level behaviour as operational conditions
change. Such capabilities represent an important step towards future
AI-native networks in which autonomous decision-making is not only
possible, but can also be deployed with operationally meaningful
assurance.
\section{Telecom Use Case: Dependency-Aware Runtime Assurance in AI-Native RAN}

The interaction between autonomous network control loops becomes a
critical challenge when multiple AI-based components influence shared
network resources. The problem is not limited to incorrect decisions
produced by individual models. A more subtle situation occurs when
individual controllers generate locally reasonable decisions according
to their own objectives and available information, while their combined
actions produce an undesirable system-level outcome. Coordination of
interacting autonomous agents has long been recognized as a fundamental
problem in multi-agent systems \cite{JenningsEtAl1998}. In O-RAN, this
problem becomes concrete: multiple independently developed xApps may
operate concurrently over overlapping network areas and pursue different
objectives, leading to direct or indirect conflicts and degraded network
performance \cite{PoleseEtAl2023,ErdolEtAl2026}.

This section illustrates this problem using an AI-native radio access
network scenario. The example considers three autonomous control loops
operating within the same RAN environment: an energy optimization
controller, a traffic prediction and capacity controller, and a service
assurance controller. Each loop has a legitimate operational objective
and may generate decisions that are individually acceptable. The
challenge arises when these decisions become coupled through shared
resources, network state, or operational constraints. The resulting
scenario is not intended to reproduce a specific existing O-RAN
deployment, but to illustrate how individually acceptable control
decisions can become incompatible when evaluated in the context of
other concurrently operating control loops.

\subsection{Scenario Description}

The considered network includes several AI-driven control functions
responsible for different operational objectives. The controllers
described below constitute an illustrative scenario rather than a
specific standardized O-RAN deployment.

The first component is an \textit{Energy Optimization Controller}. Its
objective is to reduce power consumption by adapting active network
resources to current traffic demand. During periods of low utilization,
the controller may reduce the number of active radio resources, modify
radio configurations, or place selected resources into energy-saving
modes. Such energy-aware control represents one possible optimization
function that can be implemented within increasingly programmable and
intelligent RAN architectures \cite{PoleseEtAl2023}.

The second component is a \textit{Traffic Prediction and Capacity
Controller}. It analyses observed traffic patterns and estimates future
demand. Based on these estimates, it may request additional capacity,
activate resources, or modify resource-allocation policies to reduce the
risk of future service degradation. Learning-based approaches, including
deep reinforcement learning, have been extensively investigated for
dynamic resource management and allocation in communication networks
\cite{LuongEtAl2019}.

The third component is a \textit{Service Assurance Controller}. Its
objective is to maintain service requirements expressed through
operational policies or service objectives. It monitors relevant
performance indicators, such as latency, packet loss, availability, or
other service-level metrics, and may initiate corrective actions when
observed network behaviour deviates from the intended outcome. This is
consistent with the intent lifecycle, in which intent assurance
evaluates whether the intended outcomes continue to be satisfied
\cite{RFC9315}.

Each controller can be designed and evaluated independently with respect
to its own objective. The difficulty arises when they operate
concurrently. Their decisions may affect shared resources and network
state, while each controller may observe only the information relevant
to its own control objective.

\subsection{Decision Generation and Assumption Validity}

Consider a situation where the current traffic level is low and the
Energy Optimization Controller identifies an opportunity to reduce power
consumption. It generates the following action:

\[
a_{energy} = deactivate(carrier_X)
\]

The decision is not arbitrary. It is based on a set of assumptions:

\begin{itemize}
	\item current traffic demand remains below the defined threshold,
	\item available capacity remains sufficient after the change,
	\item no other control function requires the affected resources.
\end{itemize}

At the moment of generation, the decision is valid. A runtime assurance
mechanism evaluates the action and confirms that the current network
state satisfies the required conditions.

The situation changes when the Traffic Prediction and Capacity
Controller detects an upcoming increase in demand. It generates a new
action:

\[
a_{capacity} = activate(additional\_resources)
\]

The two decisions are individually reasonable. The first one reduces
energy consumption. The second one prepares the network for increased
traffic. The problem is that they operate on the same resource domain.

The original energy optimization decision was generated under
assumptions that may no longer hold. Importantly, this does not imply
that the AI model produced an incorrect decision. Rather, the operational
context on which the decision was based changed after the decision had
been generated and evaluated.

This distinction introduces a temporal dimension to runtime assurance.
An assurance result obtained at time $t_0$ cannot necessarily be assumed
to remain valid at a later time $t_1$. If the network state or another
control action changes a condition on which the original decision
depends, the assurance result should be reconsidered before the action
is executed. In other words, assurance must consider not only whether
a decision is acceptable when evaluated, but also whether the
assumptions supporting that evaluation remain valid until execution.

\subsection{Runtime Assurance Intervention}

In the considered scenario, each controller evaluates its decisions with
respect to its own operational objective. Such local evaluation is not
sufficient to determine whether independently generated actions remain
acceptable when they affect shared network state and resources. The
assurance problem therefore concerns not only the properties of
individual decisions, but also the dependencies through which those
decisions interact.

Dependency-aware runtime assurance introduces an additional validation
step. Before execution, the assurance layer evaluates not only the action
itself, but also the assumptions and dependencies associated with that
action. This perspective is consistent with perpetual assurance, which
treats assurance as an enduring process that must incorporate new
evidence as the system and its environment evolve
\cite{WeynsEtAl2019}, and with dynamic assurance approaches that update
assurance evidence during system operation
\cite{CalinescuEtAl2018}. The mechanism considered here applies this
principle to interactions among multiple autonomous network control
loops.

For the considered case, the assurance mechanism detects:

\begin{itemize}
    \item changed traffic conditions,
    \item a dependency between the affected resources and another control loop,
    \item a conflict between energy optimization and capacity protection objectives.
\end{itemize}

The original action is therefore no longer considered valid:

\[
Valid(a_{energy},t_0)=true
\]

but:

\[
Valid(a_{energy},t_1)=false
\]

where $t_1$ represents the updated network state after the traffic
prediction event.

Instead of allowing direct execution, the assurance layer selects a
resolution strategy based on the priority of the involved control loops
and the characteristics of the conflict
(Section~\ref{sec:resolution}). In this illustrative scenario, service
continuity and capacity availability are assigned higher operational
priority than energy optimization. The assurance layer therefore
modifies the energy optimization decision:

\[
deactivate(carrier_X)
\rightarrow
reduce\_power(carrier_X)
\]

This modification preserves part of the energy optimization objective
while keeping the carrier available for the predicted increase in
capacity demand. If modification is not feasible, the assurance layer
may delay the energy optimization action until the conflicting
condition disappears or reject it if the conflict persists. If the
energy optimization action has already been executed before the new
traffic prediction becomes available, the assurance mechanism may
instead trigger rollback by reactivating $carrier_X$, provided that a
valid reversal procedure exists for that action.

The key observation is that the resolution is not arbitrary. It follows
from an explicit comparison of control-loop priorities, conflict
characteristics, action reversibility, and the available time budget.
This structured resolution process illustrates how dependency-aware
runtime assurance can reason about interactions that are not visible
when autonomous control decisions are evaluated independently.

\subsection{Research Implication}

This scenario illustrates a class of problems that emerges when
multiple autonomous control functions operate concurrently over shared
network resources and state. The challenge is not limited to the quality
of individual AI-generated decisions. A decision that is acceptable
when generated may become inappropriate for execution if the operational
conditions or dependencies on which it was evaluated subsequently
change. Evidence from O-RAN already demonstrates that independently
operating intelligent control applications can interact through shared
network parameters and objectives, producing direct or indirect
conflicts \cite{ErdolEtAl2026}. The scenario considered here extends
this observation by focusing on the temporal validity of decisions and
the assumptions on which their assurance depends.

For future autonomous networks, dependency-aware assurance should
therefore maintain awareness of three elements simultaneously:

\begin{enumerate}
    \item the current network state,
    \item the assumptions supporting individual autonomous decisions,
    \item the dependencies between interacting control loops.
\end{enumerate}

Together, these elements provide the basis for determining not only
whether a decision is acceptable when evaluated, but also whether the
conditions supporting that evaluation continue to hold until execution.
The presented scenario therefore provides a concrete basis for studying
how autonomous network control loops can be composed while maintaining
system-level operational constraints as their shared environment evolves.
\section{Initial Architecture and Formal Model for Assured Composition of AI-Native Control Loops}

The previous analysis showed that increasingly autonomous network
architectures enable multiple intelligent control functions to operate
within the same infrastructure. Evidence from O-RAN demonstrates that
independently developed control applications may interact through shared
network parameters and objectives, potentially producing direct or
indirect conflicts \cite{PoleseEtAl2023,ErdolEtAl2026}. At the same
time, research on self-adaptive systems has established that assurance
cannot necessarily be treated as a one-time activity: evidence and
assumptions may need to be reconsidered as the system and its environment
evolve \cite{WeynsEtAl2019,CalinescuEtAl2018}. Together, these
observations motivate an assurance perspective that considers not only
individual autonomous decisions, but also the operational context and
dependencies under which those decisions are generated and executed.

This section introduces an initial architecture and formal model for
dependency-aware runtime assurance of composed network control loops.
The objective is not to define a complete implementation framework, but
to establish a common representation for reasoning about controllers,
decisions, assumptions, dependencies, and changes in shared network
state. The proposed model draws on the general runtime assurance
principle of separating advanced decision-making from mechanisms
responsible for maintaining acceptable system behaviour
\cite{ShaEtAl2001,PhanEtAl2020}, while addressing the additional
problem of interactions among heterogeneous autonomous control loops.

\subsection{Architecture Overview}

Future autonomous networks are expected to include multiple intelligent
control functions operating across different network domains and
timescales. Examples include radio optimization, energy management,
traffic engineering, cloud resource orchestration, and service
assurance. Increasingly programmable network architectures provide
mechanisms for deploying such specialized control functions, while
O-RAN offers a concrete example in which multiple independently
developed xApps may operate concurrently and influence overlapping
network parameters and performance objectives
\cite{PoleseEtAl2023,ErdolEtAl2026}.

A central architectural assumption of the proposed approach is that
assurance should be separated from the internal decision-making logic of
individual autonomous controllers. An assurance decision may depend not
only on the properties of the generated action, but also on external
conditions such as the current network state, available resources,
operational constraints, and actions performed by other control
functions. Because these conditions may change after a decision has
been generated, the assurance mechanism must be able to reason about
the dependencies on which its evaluation was based.

The proposed architecture introduces a dependency-aware Runtime
Assurance Layer between autonomous control functions and the managed
network infrastructure.

\begin{figure*}[t]
	\centering
	\caption{Conceptual architecture of dependency-aware runtime assurance for AI-native network control loops.}
   \includegraphics[width=\textwidth]{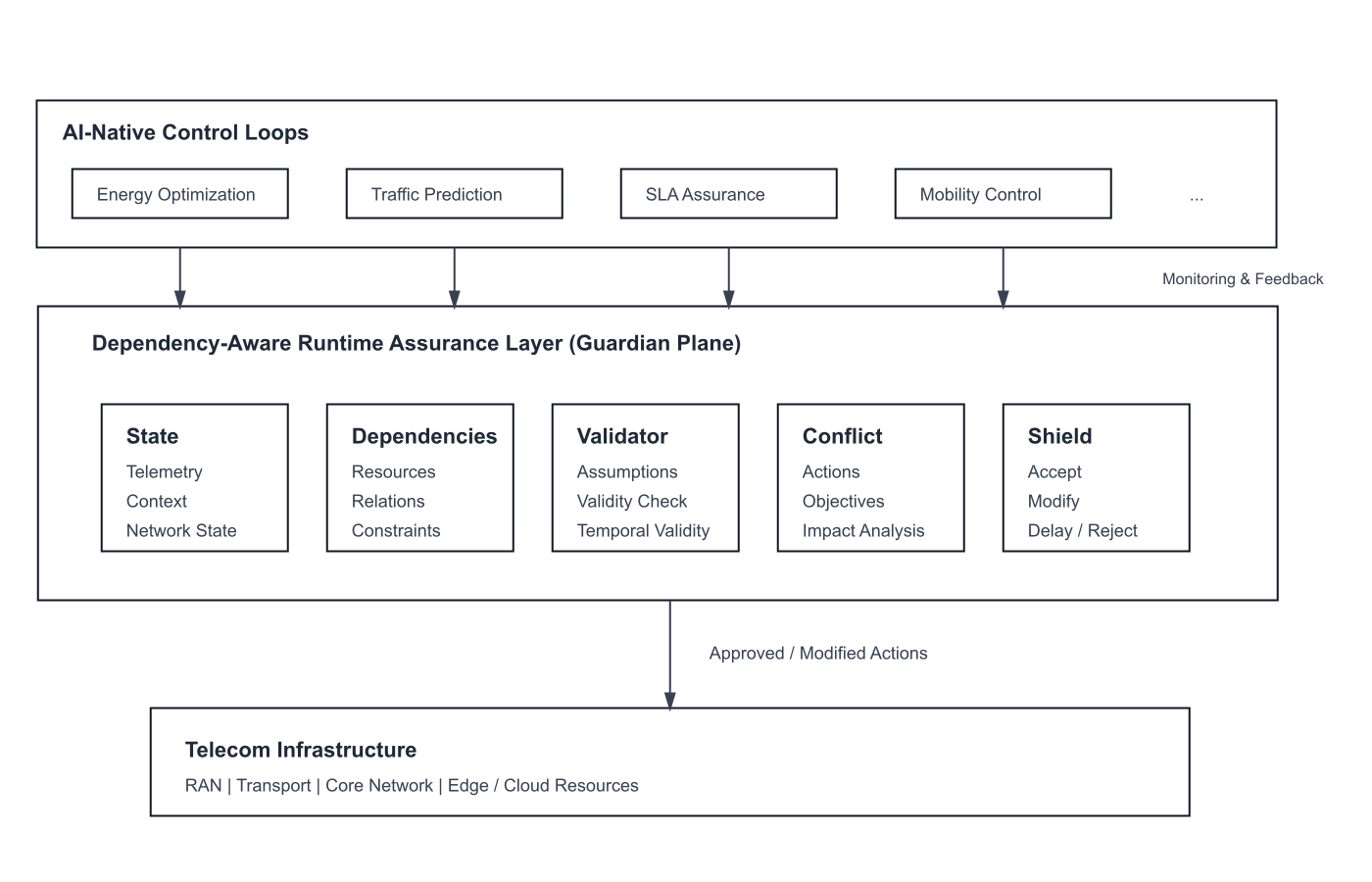}
	\label{fig:architecture}
\end{figure*}

The Runtime Assurance Layer consists of several functional elements:

\begin{itemize}
	\item \textit{State Monitoring} -- provides an up-to-date view of network conditions relevant for decision validation.
	\item \textit{Dependency Model} -- represents relationships between control loops, resources, objectives, and operational constraints.
	\item \textit{Decision Validator} -- evaluates whether an AI-generated action remains valid in the current context.
	\item \textit{Conflict Analyzer} -- identifies potentially incompatible actions generated by different control loops.
	\item \textit{Safety Enforcement Mechanism} -- allows actions to be accepted, modified, delayed, or rejected.
\end{itemize}

The role of this layer is not to replace AI-based optimization.
Autonomous controllers remain responsible for identifying optimization
opportunities and generating control decisions. The Runtime Assurance
Layer instead provides an independent mechanism for evaluating whether
those decisions remain acceptable within the current operational
context. This separation is consistent with the general principle of
runtime assurance architectures, in which advanced decision-making is
separated from mechanisms responsible for maintaining acceptable system
behaviour \cite{ShaEtAl2001,PhanEtAl2020}. The continuous
re-evaluation of decisions as operational conditions evolve is also
consistent with the perpetual assurance perspective, which treats
assurance as an enduring process throughout system operation
\cite{WeynsEtAl2019}. The proposed architecture applies these principles
to the composition of heterogeneous autonomous network control loops by
making the dependencies underlying assurance decisions explicit.

\subsection{Decision Representation}

A central assumption of the proposed model is that an AI-generated
decision cannot be represented only as an action. The action is
meaningful only together with the assumptions and dependencies under
which it was generated---a principle consistent with the uncertainty
awareness framework of Gawlikowski et al.
\cite{GawlikowskiEtAl2023} and the verified AI perspective of
Seshia et al. \cite{SeshiaEtAl2022}.

Each control loop is represented as:

\[
CL_i=(G_i,S_i,A_i,D_i)
\]

where:

\begin{itemize}
	\item $G_i$ represents the objective of the control loop,
	\item $S_i$ represents the observed system state,
	\item $A_i$ represents available actions,
	\item $D_i$ represents dependencies related to resources, constraints, and other control functions.
\end{itemize}

A generated decision is represented as:

\[
d_i=(a_i,\alpha_i,\delta_i)
\]

where:

\begin{itemize}
	\item $a_i$ is the selected action,
	\item $\alpha_i$ represents assumptions required for valid execution,
	\item $\delta_i$ represents dependencies associated with the decision.
\end{itemize}

For example, an energy optimization controller may generate an action to
deactivate a radio resource. The action itself is not sufficient to
determine whether it remains acceptable for execution. Its assurance
depends on the operational conditions under which it was evaluated,
including traffic level, available capacity, service requirements, and
possible interactions with other control functions. The decision
representation should therefore capture not only the requested action,
but also the assumptions and dependencies that determine the conditions
under which that action can be safely executed.

\begin{figure*}[t]
	\centering
	\caption{Decision representation including action, assumptions, and dependencies.}
   \includegraphics[width=\textwidth]{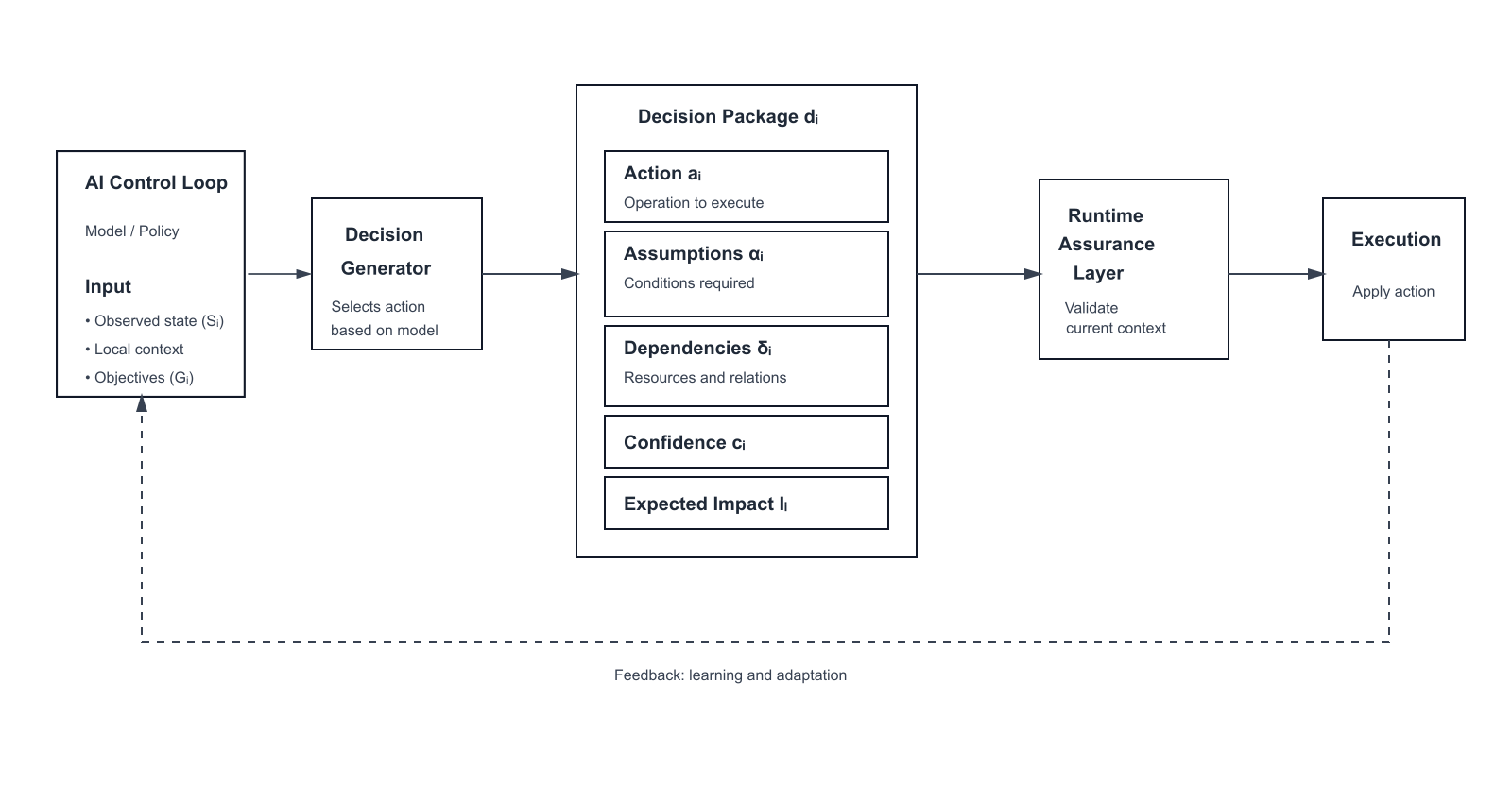}
	\label{fig:decision_representation}
\end{figure*}

\subsubsection{Acquisition of Decision Assumptions}
\label{sec:assumption_acquisition}

A fundamental question concerns the origin of $\alpha_i$. The formal
model assumes that every decision $d_i$ is associated with a set of
assumptions describing the conditions required for valid execution.
Extracting such assumptions from black-box models---deep neural
networks, LLM-based agents---is a known difficulty
\cite{GuidottiEtAl2018,LiEtAl2023}, since their internal
representations are not directly interpretable. The assurance layer
therefore does not rely on white-box access to the AI model. Instead,
$\alpha_i$ is constructed through a combination of four complementary
approaches.

\textbf{Exogenous assumption specification.}
Hard constraints---operational policies, service-level agreements,
resource limits, safety boundaries---are defined externally by the
system operator or derived from standards such as ETSI ZSM
\cite{ETSI_ZSM002} or O-RAN \cite{O-RANArchitecture2023}.
These constraints do not depend on the model's internals and are
attached to decisions based on the resources they affect and the
policies they must respect. For example, any decision that modifies a
radio resource inherits the assumptions that current traffic demand
remains below a defined threshold and that the affected resource is not
required by another control loop. This approach provides a baseline
$\alpha_i$ that is always available, regardless of model
interpretability.

\textbf{Decision-context capture.}
Rather than reverse-engineering what the model assumed internally, the
assurance layer captures the context in which the decision was
generated: the observed system state $S_i$, the input data, and the
operational conditions at decision time. The assumption $\alpha_i$ then
becomes: \emph{the conditions under which this decision was produced
remain valid}. This shifts the problem from ``what did the model
reason?'' to ``under what conditions was this decision generated, and do
those conditions still hold?''---a question that can be answered
without inspecting the model's internal logic. At runtime, the assurance
layer compares the current state $State(t)$ against the captured context
$s_i$ to detect whether the decision's premises have been invalidated.

\textbf{XAI-based approximation.}
Where some model introspection is possible, explainable AI methods can
provide partial information about which inputs influenced the decision
and within what input ranges the model is reliable. Feature attribution
methods identify the variables that most significantly shaped the
output \cite{GuidottiEtAl2018}; uncertainty quantification
techniques estimate confidence regions in which the model's predictions
are trustworthy \cite{GawlikowskiEtAl2023}. These signals do not
produce a complete specification of $\alpha_i$, but they can identify
critical input dimensions and approximate validity ranges that augment
the exogenous and context-based assumptions.

\textbf{Runtime profiling.}
By observing the behaviour of a control function across different
operational conditions over time, the assurance layer can construct an
empirical profile of the contexts in which its decisions have previously
produced acceptable or unacceptable outcomes. Observations associated
with successful execution provide evidence for conditions under which
similar decisions may be considered acceptable, while observations
followed by degradation, constraint violations, or anomalous behaviour
provide evidence of conditions requiring additional scrutiny.

Over time, this profile can provide an empirical approximation of parts
of the assumption set $\alpha_i$, complementing assumptions obtained
from explicit specifications, policies, and controller-provided
metadata. Because the profile is derived from observed behaviour rather
than formal guarantees, its conclusions should be treated as evidence
with associated uncertainty rather than as definitive proof of decision
validity.

These four approaches are complementary and can be combined. Table~\ref{tab:assumption_sources} summarizes their characteristics.

\begin{table}[t]
\centering
\caption{Approaches for acquiring decision assumptions $\alpha_i$}
\label{tab:assumption_sources}
\small
\begin{tabular}{p{0.22\columnwidth}p{0.28\columnwidth}p{0.38\columnwidth}}
\toprule
\textbf{Approach} & \textbf{Source} & \textbf{Coverage} \\
\midrule
Exogenous specification & Operator policies, standards & Hard constraints, always available \\
Context capture & Decision-time state $S_i$ & Operational conditions, model-agnostic \\
XAI approximation & Model introspection & Partial, input-dependent validity ranges \\
Runtime profiling & Observed behaviour & Empirical, improves over time \\
\bottomrule
\end{tabular}
\end{table}

The assurance layer does not require a single, complete $\alpha_i$
to be explicitly provided by the autonomous controller. Instead, it
operates on the assumption information that is available from sources
such as external constraints, captured operational context,
controller-provided confidence information, or empirical runtime
profiles. Consequently, $\alpha_i$ is treated as an incrementally
constructed and potentially incomplete set.

When the available assumption information is incomplete, the assurance
layer adopts a conservative interpretation of decision validity. For
example, it may initially assume that a decision remains acceptable only
while the relevant operational conditions observed at generation time
remain unchanged. Such restrictions may subsequently be relaxed as
additional evidence about the decision dependencies becomes available.

This design allows the assurance mechanism to operate even when the
internal decision logic of the controller is not directly accessible.
This is particularly relevant for black-box learning models, whose
internal reasoning may be difficult to interpret
\cite{GuidottiEtAl2018}, and for LLM-based agents considered for network
operations and management, where reliability, explainability, and
integration with operational systems remain important challenges
\cite{LiuEtAl2025LLMNetOps}.

\subsection{Runtime Validation of Autonomous Decisions}

In traditional automated systems, decision validity can often be
evaluated against predefined rules and explicitly represented system
conditions. Autonomous network control requires a broader perspective:
the operational conditions and assumptions under which a decision was
evaluated may change after the decision has been generated. Research on
self-adaptive systems provides an important foundation for addressing
such changes by combining runtime modelling and verification with
dynamically maintained assurance evidence
\cite{CalinescuEtAl2018}. The perpetual assurance perspective similarly
emphasizes that assurance must continuously incorporate new evidence as
the system and its environment evolve
\cite{WeynsEtAl2019}.

For dependency-aware runtime assurance, this motivates treating decision
validity as a time-dependent property rather than as a permanent result
of a one-time validation.

The validity of a decision at time $t$ can be expressed as:

\[
Valid(d_i,t)=
State(t)\models \alpha_i
\land
Dependencies(t)\approx\delta_i
\]

A decision remains valid only when:

\begin{enumerate}
	\item the current network state satisfies the assumptions associated with the decision,
	\item relevant dependencies remain unchanged or within acceptable limits.
\end{enumerate}

This introduces a temporal dimension. A decision accepted at one moment
may become invalid later without any failure of the underlying AI model.

\[
Valid(d_i,t_0)=true
\]

does not imply:

\[
Valid(d_i,t_1)=true
\]

when the operational conditions on which the original assurance decision
depended have changed. This temporal invalidation is precisely the
scenario illustrated in the use case of Section~6. In the proposed
model, the problem is therefore not only to evaluate a decision at the
time it is generated, but also to identify the dependencies on which
that evaluation relies and determine whether subsequent changes to those
dependencies require the decision to be re-evaluated.

This motivates a continuous assurance process rather than treating
validation as a one-time result. Such a perspective is consistent with
the perpetual assurance principle, which treats assurance as an
enduring process that incorporates new evidence as the system and its
environment evolve \cite{WeynsEtAl2019}.

\subsection{Composition of Multiple Control Loops}

The main challenge addressed by assured composition is the interaction
between autonomous control loops. A closely related problem is well
recognized in multi-agent reinforcement learning, where non-stationarity
is considered a fundamental challenge: as multiple agents act and adapt
within a shared environment, the environment observed by an individual
agent changes as the policies and actions of other agents evolve
\cite{GronauerDiepold2022}.

For two control loops:

\[
CL_i \oplus CL_j
\]

the combined behaviour depends not only on their individual decisions
but also on the relationships between those decisions and their
dependencies.

A direct conflict may occur when:

\[
Conflict(a_i,a_j)=true.
\]

In addition, the condition:

\[
D_i \cap D_j \neq \emptyset
\]

indicates that the two decisions share at least one dependency and may
therefore interact through common network state, resources, or
operational constraints. Such interactions are already observable in
O-RAN, where independently developed xApps may influence overlapping
network parameters and performance objectives, resulting in direct or
indirect conflicts \cite{ErdolEtAl2026}.

The existence of shared dependencies does not automatically imply a
conflict. Many network control functions naturally operate on common
resources or depend on common elements of network state. Instead,
shared dependencies identify relationships that may require additional
assurance when the corresponding control decisions are executed
concurrently. The challenge is therefore to determine whether changes
to a shared dependency affect the validity of one or more decisions and
whether their simultaneous execution remains compatible with
system-level operational objectives.

\begin{figure*}[t]
	\centering
	\caption{Runtime assurance process for interacting AI-native control loops.}
    \includegraphics[width=\textwidth]{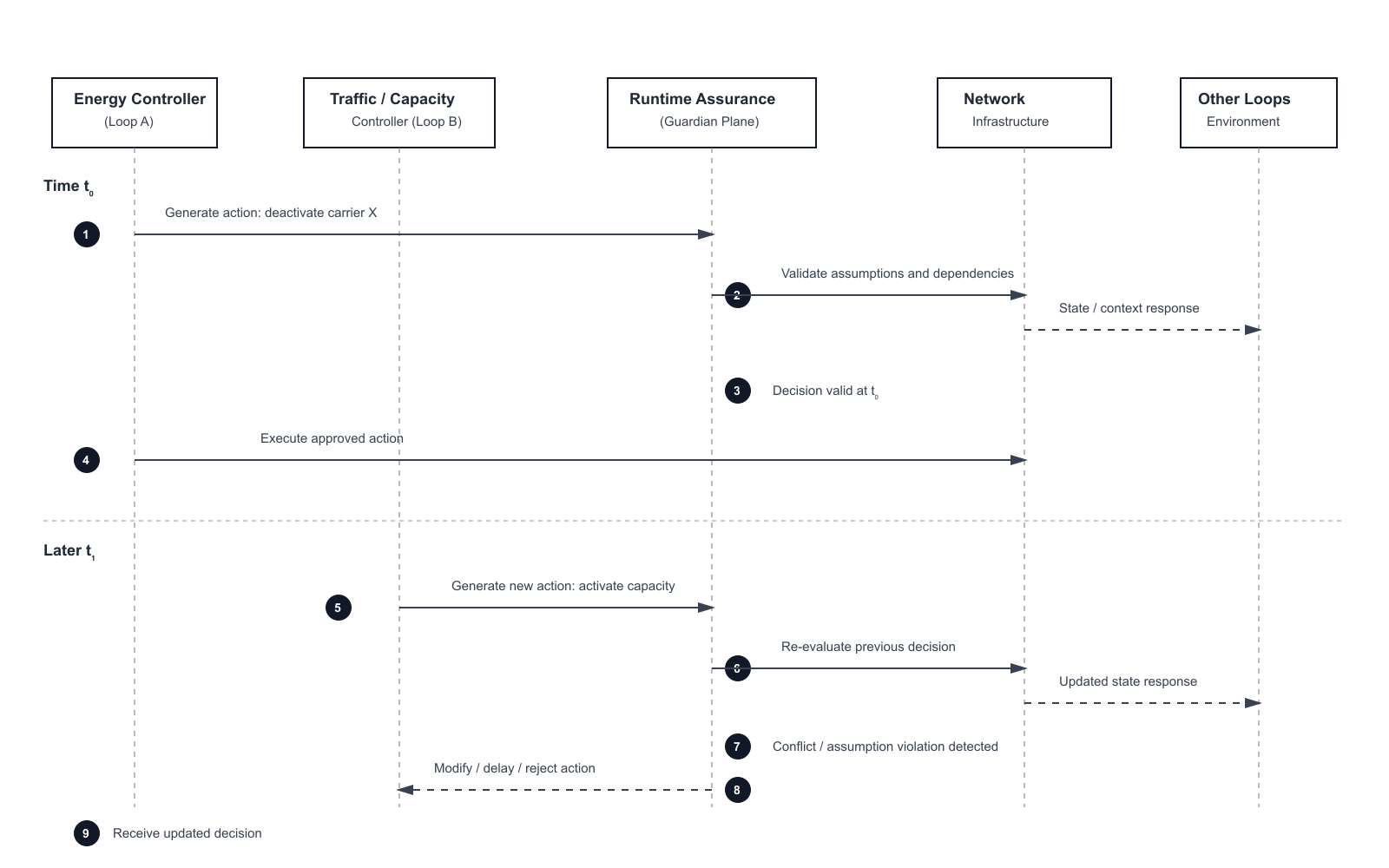}
	\label{fig:sequence_diagram}
\end{figure*}

\subsection{Resolution Strategies and Priority Mechanisms}
\label{sec:resolution}

When the assurance layer detects that $Valid(d_i, t) = false$ or that
$Conflict(a_i, a_j) = true$, it must determine an appropriate response.
The Safety Enforcement Mechanism can accept, modify, delay, or reject an
action---and, in some cases, initiate rollback of a previously executed
decision. The choice among these outcomes is not arbitrary; it depends on
the nature of the conflict, the operational context, and the relative
priority of the involved control loops. This subsection discusses how
resolution decisions are made.

\textbf{Decision outcome taxonomy.}
The assurance layer distinguishes five outcomes, ordered by increasing
intervention severity:

\begin{enumerate}
	\item \textit{Accept}: the action is valid and compatible with other
	active decisions; it is executed as generated.
	\item \textit{Modify}: the action is adjusted to eliminate the
	conflict while preserving the controller's objective as closely as
	possible (e.g., $deactivate(carrier_X) \rightarrow reduce\_power(carrier_X)$,
	as illustrated in Section~6).
	\item \textit{Delay}: the action is postponed until the conflicting
	condition resolves (e.g., until the traffic prediction window expires
	or another controller releases the resource).
	\item \textit{Reject}: the action is blocked entirely because no safe
	modification or delay can resolve the conflict within the available
	time budget.
	\item \textit{Rollback}: a previously executed action is reversed
	because its assumptions were invalidated after execution, and the
	resulting state poses a risk to service continuity or other control
	loops.
\end{enumerate}

The selection among these outcomes depends on two factors: the
\emph{priority} of the involved control loops and the
\emph{characteristics} of the conflict.

\textbf{Priority assignment.}
Each control loop $CL_i$ is assigned a priority that reflects the
operational importance of its current decision. Priority is not a
property of the AI model itself, but of the operational role, policy
context, and timing constraints associated with the control action.
Three dimensions determine priority:

\begin{itemize}
    \item \textit{Service criticality}: controllers responsible for
    service continuity, safety, or compliance with service-level
    requirements may take precedence over optimization-oriented
    controllers. For example, a service assurance action may be assigned
    higher priority than an energy optimization action when the latter
    would risk service degradation.

    \item \textit{Policy precedence}: operator-defined policies and
    intent specifications may establish explicit precedence relationships
    between operational objectives \cite{RFC9315}. Such policies may
    encode domain-specific rules (e.g., ``radio resource availability
    takes precedence over energy saving during peak hours'') or
    service-tier hierarchies.

    \item \textit{Temporal urgency}: decisions whose usefulness or
    validity is constrained by a shorter execution window may be assigned
    higher priority when competing actions cannot be executed
    simultaneously. Delaying such a decision may reduce its effectiveness
    or render it obsolete. In O-RAN, for example, Near-RT RIC control
    loops typically operate at timescales from approximately 10~ms to
    1~s, whereas Non-RT control loops operate at timescales of 1~s or
    longer \cite{PoleseEtAl2023}.
\end{itemize}

Priority is represented as an attribute of $CL_i$, integrated into the
control loop model:
\[
CL_i = (G_i, S_i, A_i, D_i, P_i)
\]
where $P_i$ denotes the priority level. When a conflict is detected
between $CL_i$ and $CL_j$, the assurance layer compares $P_i$ and $P_j$.
The lower-priority controller's action is modified, delayed, or
rejected; the higher-priority action proceeds.

\textbf{Conflict characteristics and resolution criteria.}
Priority alone does not determine the outcome. The assurance layer also
evaluates:

\begin{itemize}
	\item \textit{Reversibility}: if the action can be undone without
	significant cost (e.g., reactivating a carrier), modification or
	delay is preferred. If the action is irreversible or expensive to
	reverse (e.g., migrating a service instance), rejection or more
	conservative modification is warranted.
	\item \textit{Severity of impact}: what happens if the conflict is
	allowed? If the impact is limited to reduced optimization
	performance, modification or delay may suffice. If the impact
	threatens service availability or violates safety constraints,
	rejection or rollback is necessary.
	\item \textit{Time budget}: within tight latency constraints
	(Section~\ref{sec:latency_budget}), only fast resolution strategies
	are feasible. If no resolution can be computed within the available
	time, the assurance layer applies a conservative default: reject the
	lower-priority action and allow the higher-priority action to
	proceed.
\end{itemize}

\textbf{Rollback strategy.}
Rollback is the most disruptive resolution and is reserved for cases
where a previously executed action's assumptions were invalidated after
execution---i.e., $Valid(d_i, t_0) = true$ but $Valid(d_i, t_1) = false$
\emph{after} $d_i$ has already been applied. Rollback requires that the
assurance layer maintain, for each executed action, a record of the
previous state and a reversal procedure. Not all actions are
reversible: some network changes (e.g., configuration commits, service
migrations) may require compensating actions rather than direct
reversal. The assurance layer should therefore classify actions by
reversibility at decision time and maintain compensating action plans
for non-reversible operations.

\textbf{Unilateral vs.\ negotiated resolution.}
The architecture supports two resolution modes. In \emph{unilateral}
mode, the assurance layer independently determines the resolution based
on decision priorities, conflict characteristics, operational policies,
and the available time budget. This mode is particularly suitable when
the resolution must be completed within a short control-loop deadline
and iterative coordination between controllers would introduce
unacceptable latency.

In \emph{negotiated} mode, the assurance layer informs the conflicting
controllers about the detected conflict and allows them to propose or
adjust their decisions within a bounded coordination window. This mode
builds on the general principle of coordination among autonomous agents,
where interacting agents may exchange information and adapt their
actions to achieve compatible outcomes \cite{JenningsEtAl1998}.
Negotiated resolution is applicable when the available time budget is
sufficient to accommodate the additional communication and decision
iterations. It may therefore be particularly relevant to slower control
loops, including Non-RT RIC and orchestration functions, but is not
defined exclusively by a particular control tier.

The choice between unilateral and negotiated resolution is itself
policy-driven and may depend on conflict severity, decision priority,
operational context, and the time available for resolution.

Table~\ref{tab:resolution_strategies} summarizes the resolution
strategies and their applicability.

\begin{table}[t]
\centering
\caption{Resolution strategies and applicability criteria}
\label{tab:resolution_strategies}
\small
\begin{tabular}{p{0.18\columnwidth}p{0.32\columnwidth}p{0.38\columnwidth}}
\toprule
\textbf{Outcome} & \textbf{When applied} & \textbf{Typical scenario} \\
\midrule
Accept & No conflict detected & Normal operation \\
Modify & Conflict solvable by adjustment & Reduce scope of action to avoid resource overlap \\
Delay & Conflict is transient & Postpone energy saving until traffic peak passes \\
Reject & No safe resolution within time budget & Irreversible action threatens service \\
Rollback & Assumptions invalidated post-execution & Previously deactivated resource needed by higher-priority loop \\
\bottomrule
\end{tabular}
\end{table}

The resolution framework outlined here is necessarily initial. The
definition of precise priority policies, the design of efficient
reversal procedures, and the evaluation of negotiated versus unilateral
resolution under realistic workloads are identified as key research
directions in Section~9.

\subsection{Latency Budget and Scalability Considerations}
\label{sec:latency_budget}

A critical concern for any assurance mechanism operating in
telecommunication networks is the latency overhead it introduces.
Autonomous control functions may operate at substantially different
timescales. In O-RAN, for example, Near-RT RIC control loops typically
operate at timescales between approximately 10~ms and 1~s, whereas
Non-RT RIC control operates at timescales of 1~s or longer
\cite{PoleseEtAl2023}. Other management and orchestration functions
may operate at substantially longer timescales, depending on the
specific operational process.

These control-loop timescales should not be interpreted directly as
latency budgets available to the Runtime Assurance Layer. The assurance
mechanism shares the available control-cycle time with decision
generation, communication, enforcement, and other processing required
before an action becomes effective. For a control loop with an
end-to-end time constraint $T_{\mathrm{control}}$, the basic timing
condition can therefore be expressed as:

\[
T_{\mathrm{decision}}
+
T_{\mathrm{assurance}}
+
T_{\mathrm{enforcement}}
\leq
T_{\mathrm{control}}.
\]

Consequently, the maximum time available for assurance is bounded by:

\[
T_{\mathrm{assurance}}
\leq
T_{\mathrm{control}}
-
T_{\mathrm{decision}}
-
T_{\mathrm{enforcement}}.
\]

Introducing a Runtime Assurance Layer that analyses dependencies,
detects conflicts, validates assumptions, and selects appropriate
enforcement actions inevitably consumes part of this budget. The
feasibility of the proposed architecture therefore depends not only on
the nominal timescale of the corresponding control loop, but on whether
the required assurance operations can be completed within the remaining
end-to-end latency budget.

\textbf{Tiered time budgets.}
The assurance layer is not intended to operate uniformly across all
control loops. Assurance requirements and available time budgets depend
on the timescale of the corresponding control function. In O-RAN, for
example, Near-RT RIC control loops typically operate at timescales
between approximately 10~ms and 1~s, whereas Non-RT RIC control
operates at timescales of 1~s or longer \cite{PoleseEtAl2023}.
Consequently, fast control loops require lightweight assurance
operations, while slower loops may permit more comprehensive dependency
and conflict analysis. The actual assurance budget is only a fraction
of the control-loop timescale, as decision generation and enforcement
consume part of the available time.

\textbf{Bounded-latency assurance operations.}
Different assurance functions have different computational costs.
Validation of explicitly represented assumptions may involve simple
comparisons with current network state, whereas conflict analysis may
require examination of relationships between multiple decisions and
control loops. For time-constrained loops, the architecture therefore
distinguishes a fast synchronous path from deeper analysis that can be
performed asynchronously. The synchronous path contains checks required
before execution, while additional analysis can update information used
in subsequent assurance cycles.

\textbf{Asynchronous assurance.}
The resulting hybrid model combines synchronous validation with
asynchronous analysis. This approach allows operations that are not
required on the critical execution path to proceed independently while
their results update the dependency model and inform future decisions.
Such continuous revision of assurance evidence is consistent with the
perpetual assurance perspective \cite{WeynsEtAl2019}.

\textbf{Scalability considerations.}
The computational cost of assurance depends on the number of active
control loops and the dependencies between them. A dependency graph
allows analysis to focus on control loops and resources related to the
current decision rather than considering all possible interactions.
This becomes increasingly important as the number of controllers grows,
because exhaustive pairwise analysis of $N$ control loops may require
up to

\[
\frac{N(N-1)}{2}
\]

comparisons. Partitioning control loops by domain or resource scope may
further reduce the number of interactions requiring analysis.

\textbf{Trade-off between assurance depth and latency.}
The architecture therefore exposes a trade-off between assurance depth
and processing latency. Fast control loops may support only a limited
set of checks before execution, whereas slower loops may permit deeper
dependency and conflict analysis. The appropriate assurance depth can
depend on the control-loop requirements, operational context, and
available time budget.

The actual latency and scalability characteristics of the proposed
architecture must be established experimentally. Open and programmable
mobile-network platforms provide environments for evaluating intelligent
control functions \cite{BonatiEtAl2020,PoleseEtAl2023}, while Network
Digital Twins provide complementary capabilities for controlled
evaluation of network behaviour
\cite{MasaracchiaEtAl2022,VilaEtAl2023}. Experimental evaluation should
determine the overhead of individual assurance operations and establish
which assurance mechanisms are feasible at different control-loop
timescales.

\subsection{Research Perspective}

The presented abstraction defines the elements required to study assured
composition of AI-native control loops:

\begin{itemize}
	\item autonomous entities generating decisions,
	\item assumptions describing decision validity,
	\item dependencies linking control loops and resources,
	\item runtime mechanisms evaluating compatibility.
\end{itemize}

This model shifts the research focus from improving individual AI
decisions towards understanding how multiple autonomous decisions can
coexist within a shared infrastructure. The proposed architecture and
formal representation provide a basis for further investigation of
dependency modelling, conflict detection mechanisms, runtime
verification approaches, and experimental evaluation methods for future
autonomous telecom networks.

\section{Conclusions}

The evolution towards AI-native networks represents a significant change
in how future telecommunication infrastructures may be operated and
controlled. AI is evolving from a supporting tool for network analytics
towards an increasingly integral element of network control and
management architectures
\cite{LetaiefEtAl2024,ZhangEtAl2025AINative}. In parallel, advances in
zero-touch management, programmable network control, Network Digital
Twins, and autonomous decision-making provide important foundations for
increasing levels of network autonomy
\cite{CoronadoEtAl2022,PoleseEtAl2023,RazaEtAl2025}.

Increasing the intelligence of individual control functions, however,
does not by itself ensure acceptable system-level behaviour when
multiple autonomous functions operate concurrently. Programmable
architectures such as O-RAN already demonstrate that independently
developed control applications may interact through shared network
parameters and performance objectives, producing direct or indirect
conflicts \cite{PoleseEtAl2023,ErdolEtAl2026}. As network control
becomes increasingly distributed across functions operating at
different timescales and pursuing different objectives, assurance must
therefore consider not only individual decisions but also the
interactions between them and the operational conditions on which their
validity depends.

This paper reviewed the evolution from traditional network management
towards AI-native autonomous control and examined relevant developments
in intelligent controllers, Network Digital Twins, AIOps, autonomous
agents, trustworthy AI, runtime assurance, and multi-agent coordination.
The literature provides substantial foundations for this transition.
Network Digital Twins support modelling, simulation, prediction, and
evaluation of network behaviour
\cite{MasaracchiaEtAl2022,RazaEtAl2025}; AIOps and learning-based
networking provide increasingly sophisticated operational intelligence
and optimization
\cite{ZhangEtAl2019,ZhangEtAl2025AIOps}; trustworthy AI addresses
properties including robustness, reliability, transparency, and
explainability \cite{LiEtAl2023,NISTAIRMF2023}; and runtime assurance
provides mechanisms for maintaining acceptable operation in the
presence of complex or learning-enabled controllers, including
distributed multi-agent systems
\cite{PhanEtAl2020,MehmoodEtAl2021}.

Taken together, however, these capabilities expose a further assurance
problem at the level of composed network control. A decision that is
acceptable when generated may depend on assumptions about network state,
resources, policies, or other control functions that subsequently
change. The relevant question is therefore not only whether an
individual autonomous decision is acceptable at a particular instant,
but whether the conditions supporting that assurance decision remain
valid until execution and while other autonomous control loops modify
the shared operational environment.

This survey identifies dependency-aware runtime assurance as a
promising research direction for addressing this problem. Such an
approach would explicitly represent the assumptions and dependencies
associated with autonomous decisions, monitor relevant changes in the
operational environment, trigger re-evaluation when those dependencies
change, and provide mechanisms for resolving interactions between
concurrent control actions. The objective is not to constrain autonomous
optimization unnecessarily, but to enable heterogeneous AI-native
control loops to be composed while maintaining acceptable system-level
behaviour under dynamically changing network conditions.

The resulting research question can be stated as follows:

\begin{quote}
How can multiple AI-native network control loops be composed and
assured at runtime when the validity of their decisions depends on
shared and dynamically changing network state?
\end{quote}

Addressing this question requires further work on formal dependency
representations, runtime decision validation, conflict resolution,
latency-aware assurance mechanisms, and reproducible experimental
evaluation. Together, these elements provide a research agenda towards
assured composition of AI-native network control loops and, ultimately,
towards autonomous telecommunication networks in which increasing
decision-making capability is accompanied by operationally meaningful
assurance.

\bibliographystyle{IEEEtran}
\bibliography{sample}

\end{document}